\documentclass[reprint,superscriptaddress,floatfix,longbibliography,tightenlines,nobibnotes,nofootinbib,amsmath,amssymb,aps,pra]{revtex4-2}
\usepackage{graphicx}
\usepackage{color}
\usepackage{bm}
\usepackage{dcolumn}
\usepackage{textcomp}
\usepackage{hyperref}
\hypersetup{colorlinks=true,urlcolor= blue,citecolor=blue,linkcolor= blue,bookmarks=true,bookmarksopen=false}
\usepackage{verbatim}
\usepackage{blindtext}
\usepackage{natbib}
\setcitestyle{numbers,square}
\usepackage{makecell}

\begin{document}

\title{Influence of interface-induced valley-Zeeman and spin-orbit couplings\\ on transport in graphene-on-WSe$_{2}$ heterostructures}
\author{M. Zubair}
\email{muhammad.zubair@mail.concordia.ca}
\affiliation{Department of Physics, Concordia University, 7141 Sherbrooke Ouest, Montreal, Quebec H4B 1R6, Canada}
\author{P. Vasilopoulos }
\email{p.vasilopoulos@concordia.ca}
\affiliation{Department of Physics, Concordia University, 7141 Sherbrooke Ouest, Montreal, Quebec H4B 1R6, Canada}
\author{M. Tahir}
\email{tahir@colostate.edu}
\affiliation{Department of Physics, Colorado State University, Fort Collins, CO 80523, USA}

\begin{abstract}

We investigate the electronic dispersion and transport properties of graphene/WSe$_{2}$ heterostructures in the presence of a proximity induced spin-orbit coupling (SOC) 
 using a low-energy  Hamiltonian, with different types of symmetry breaking terms, obtained from a four-band, first and second nearest-neighbour tight-binding (TB) one. The competition between different perturbation terms leads to inverted SOC bands. Further, we study the effect of symmetry breaking terms on ac and dc transport by evaluating the corresponding conductivities within linear response theory. The scattering-independent part of the valley-Hall conductivity, as a function of the Fermi energy $E_{F}$, is mostly negative in the ranges $-\lambda_{R}\leqslant E_{F}$ and $E_{F}\geqslant\lambda_{R}$ when the strength $\lambda_{R}$ of the Rashba SOC increases except for a very narrow region around $E_{F}=0$ in which it peaks sharply upward. The scattering-dependent diffusive conductivity increases linearly with electron density,  is directly proportional to $\lambda_{R}$ 
 in the low- and high-density regimes, but weakens for $\lambda_{R}=0$. We investigate the optical response in the presence of a SOC-tunable band gap for variable $E_{F}$. An interesting feature of this SOC tuning is that it can 
 be used to switch on and off the Drude-type intraband response. Furthermore, the ac conductivity exhibits  interband responses due to 
 the Rashba SOC. We also show that the valley-Hall conductivity changes sign when $E_F$ is comparable to $\lambda_R$ and vanishes at higher values of $E_F$. It also 
 exhibits a strong  dependence on temperature and  a considerable structure as a function of the frequency.

\end{abstract}

\maketitle
	
\section{Introduction}

Two-dimensional (2D) materials have  become a hot topic in solid state physics, especially since the discovery of graphene,  both theoretically and experimentally because of their prominent mechanical, optical, electrical and magnetic properties \cite{r1}. Recently graphene has attracted a lot of attention 
in the field of spintronics due to its large electronic mobility, low spin-orbit coupling (SOC), negligible hyperfine interaction and gate tunability \cite{r2}. For a clear example, it has been proven that graphene exhibits a very long spin relaxation length even  at room temperature \cite{r3,r4}. Due to the weak SOC though, it is not a suitable candidate for the observation of important spin-dependent phenomena including the spin-Hall effect \cite{r5} and anomalous Hall effect \cite{r6}.

To render graphene useful in spintronics, 
several experimental groups used  different techniques to tailor  the SOC strength in it through coupling with foreign atoms or materials \cite{r7,r8,r9,r10,r11,r12,r13}, such as graphene hydrogenation \cite{r14,r15} or fluorination \cite{r16} as well as heavy adatom decoration \cite{r17,r18}. However, these approaches not only reduce the transport quality, but also make it difficult to reproduce \cite{r14,r15} and detect \cite{r16,r17,r18} the induced SOC. To overcome these difficulties, graphene is recently grown on different novel 2D materials, which are ideal candidates to induce SOC via proximity effects \cite{rr19, rr20, rr21, rr22,rr23,rr24,rr25}. Hexagonal boron nitride (BN) has a weak SOC, and therefore, is not a suitable substrate for the proximity effect \cite{r19}. The family of 2D transition metal dichalcogenides (TMDCs) are the next best candidates, which have large direct band gaps and giant intrinsic SOC \cite{r20,r21}. In this respect, graphene on TMDCs has been investigated for transport \cite{r22,r23, nnn1} as well as intriguing technological applications, including  field-effect tunnelling transistors (FETTs), radio-frequency oscillators, and efficient phototransistors \cite{r24,r25,r26,r27,rr26,rr27,rr28}.  Also,  
the proximity-induced SOC in graphene/TMDCs heterostructures has  recently been shown to depend \cite{nnn2, nnn3} on the twist angle between  the lattice of graphene and that of the TMDC.

 In addition, it has been found in room-temperature experimental studies of the  spin-Hall effect that few-layer WS$_{2}$ induces a large SOC in graphene, about $17$ meV \cite{r28} as compared to the very weak one in pristine graphene \cite{r29}. Also, it has been unambiguously demonstrated experimentally that a room-temperature spin-Hall effect in graphene is induced by MoS$_{2}$ proximity \cite{r30}. Moreover, when graphene is placed on a multilayer WS$_{2}$ substrate, an additional valley-Zeeman SOC, due to the broken sublattice symmetry, along with the Rashba SOC have been predicted theoretically and observed experimentally  \cite{rr25, r31, rr31,rr32}. This SOC induces a spin splitting of degenerate bands, with out-of-plane spin polarization at the $K$ and $K^{\prime}$ points, and an opposite spin splitting in different valleys. 
%
Analogous to the Zeeman splitting, the SOC is termed 
 valley-Zeeman because the effective Zeeman fields are valley-dependent. It is the dominant SOC in TMDCs and is also predicted to be induced in graphene on TMDCs \cite{rr25, r31, rr31,rr32}. To our knowledge though, apart from some spin-transport studies \cite{st} and two experimental magneto-transport studies \cite{mt}, neither ac and dc scattering-dependent charge transport nor the simultaneous effect of valley-Zeeman and Rashba SOCs have been theoretically studied in graphene on TMDCs. 

In this work we study in detail the effect of the valley-Zeeman and Rashba-type SOCs on ac and dc transport in graphene/WSe$_{2}$ heterostructures. There results 
a  mexican hat dispersion \cite{rr33} contrary to  other family memebers of TMDCs , e.g., MoS$_{2}$, WS$_{2}$ etc. \cite{rr34}. Such a  dispersion leads to more features in the  optical conductivity when the Fermi level moves between the minimum and maximum of the mexican  hat.  Also, we compare our results with those for pristine graphene.

In Sec. II we specify the Hamiltonian and obtain the 
eigenvalues and eigenfunctions in the presence of symmetry breaking terms. In Sec. III we present  general expressions for the conductivities and provide numerical results. Conclusions and a summary follow in Sec. IV.
\begin{figure}[t]
\centering
\includegraphics[width=8cm, height=6cm]{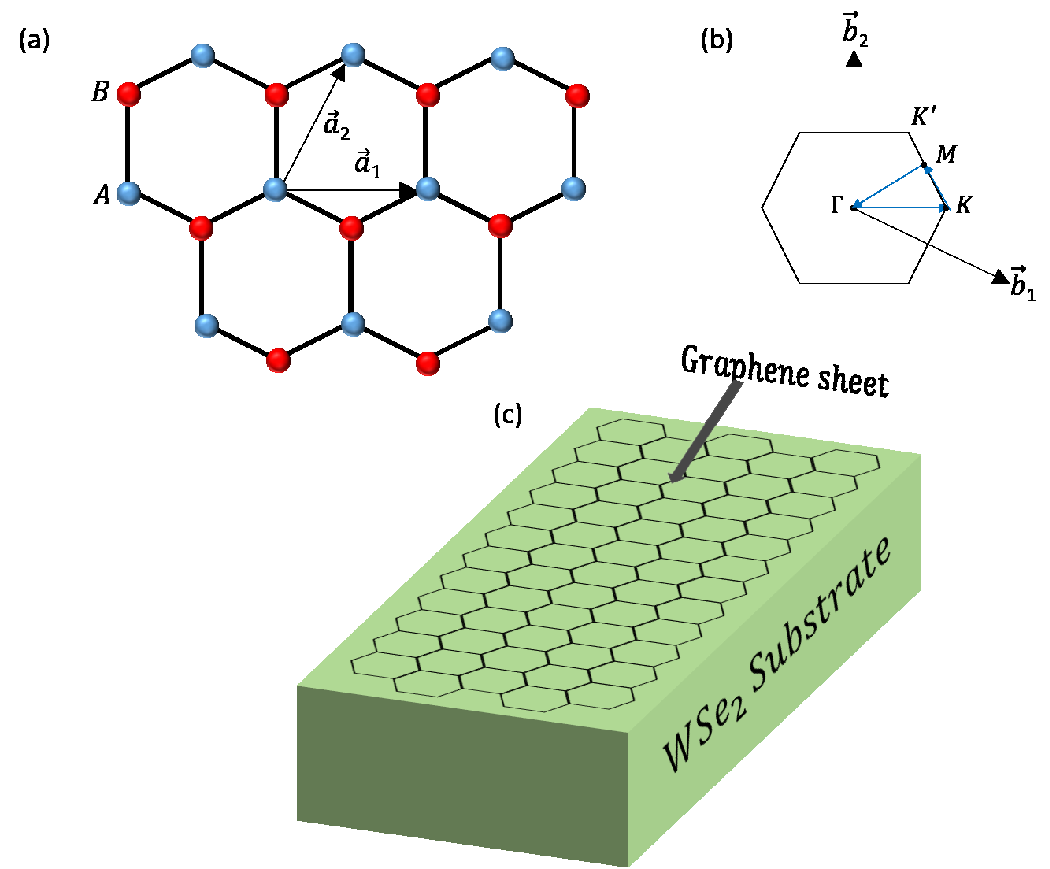}
\caption{(a) Real-space graphene with $\vec{a}_{1}$ and $\vec{a}_{2}$ the primitive lattice vectors. (b) Graphene's first Brillouin zone and high symmetry points $\Gamma$, $K$, and $M$  in reciprocal space. 
Its primitive lattice vectors are $\vec{b}_{1}$ and $\vec{b}_{2}$. (c) Schematic representation of graphene on a WSe$_{2}$ substrate. }
\label{f1}
\end{figure}
\section{Formulation}

Graphene is a 2D, one-atom thick planar sheet of bonded carbon atoms densely packed in a honeycomb structure as shown in Fig. \ref{f1} $(a)$. The lattice structure can be viewed as a triangular lattice with  two sites $A$ (red filled spheres) and $B$ (blue filled spheres) per unit cell. The arrows indicate the primitive lattice vectors $\vec a_{1}= a (1, 0)$ and $\vec a_{2}= a(1/2, \sqrt{3}/2)$, with $a$ the triangular lattice constant of the structure,  and span the graphene lattice. Further, $\vec a_{1}$ and $\vec a_{2}$ generate the  reciprocal lattice vectors of the Brillouin zone, cf.  Fig. \ref{f1} $(b)$, given by $\vec b_{1}=4\pi/\sqrt{3} a  (\sqrt{3}/2, - 1/2)$ and $\vec b_{2}=4\pi/\sqrt{3} a (0, 1)$. From the explicit expressions of  $\vec b_{1}$ and $\vec b_{2}$ we find the two inequivalent Dirac points (valleys) given by  $\vec K =  (4 \pi/3a) (1, 0)$ and $\vec K^{\prime}=(4 \pi/3 a) (1/2, \sqrt{3}/2)$. 
\begin{figure}[t]
\centering
\includegraphics[width=9.5cm, height=9cm]{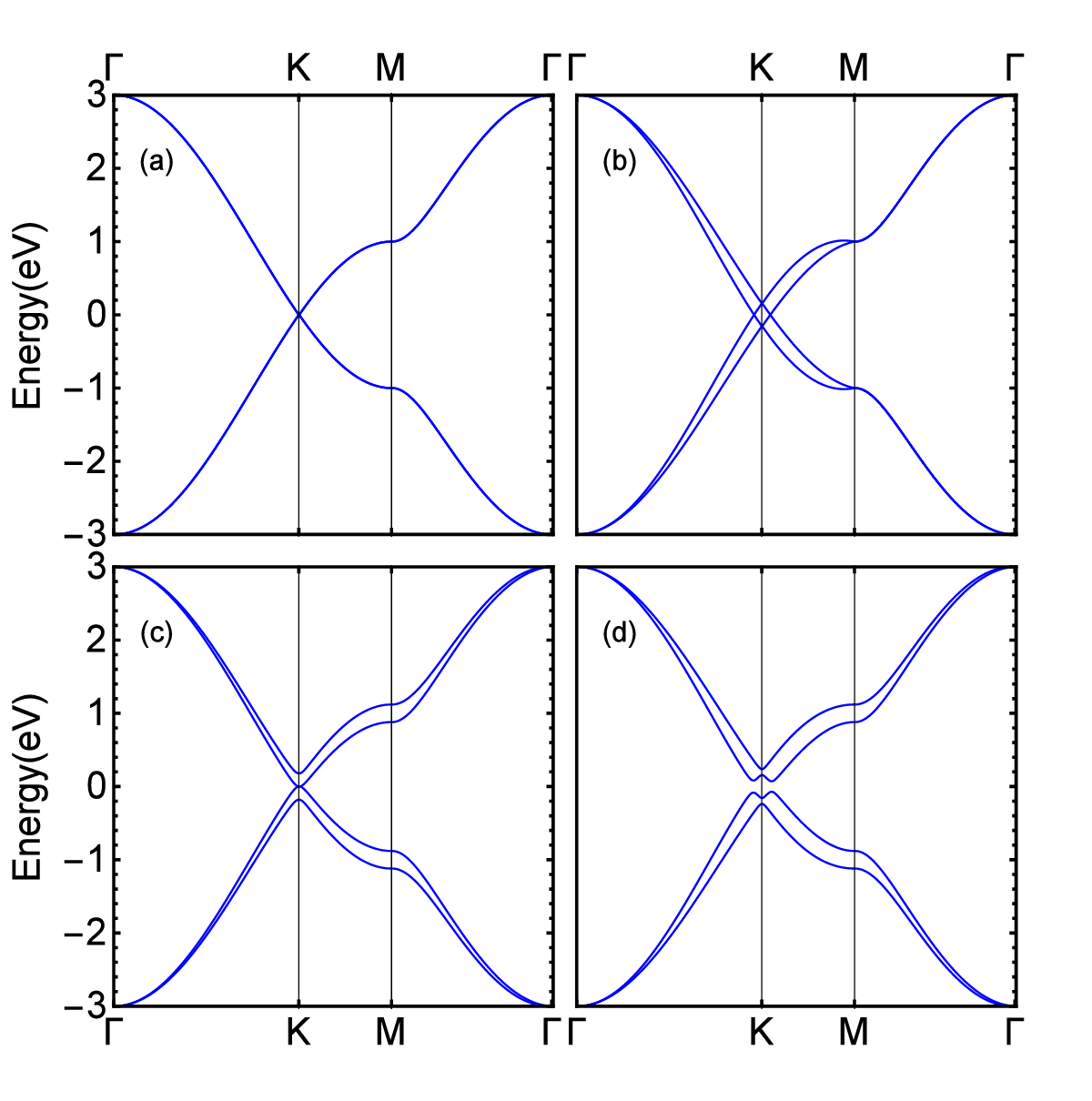}
\caption{ Energy dispersion in a graphene/WSe$_{2}$ heterostructure  using the TB model (\ref{tb1}) along the path $-M \rightarrow -K \rightarrow \Gamma \rightarrow K \rightarrow M$ for (a) $\lambda_{c_{i}}, \lambda_{R}=0$, (b) $\lambda_{c_{i}} \neq 0, \lambda_{R}=0$, (c) $\lambda_{c_{i}} = 0, \lambda_{R} \neq 0$, and (d) $\lambda_{c_{i}}, \lambda_{R} \neq 0$  .}
\label{f2}
\end{figure}

The monolayer graphene system is described by the four-band, second nearest-neighbour tight-binding  (TB) Hamiltonian \cite{rr23, rr33, r32}
\begin{eqnarray}
\notag
H & = & \sum_{\langle i,j \rangle,\alpha} t c_{i\alpha}^{\dagger} c_{j\alpha} +\sum_{i \alpha} \Delta \eta_{c_{i}} c_{i\alpha}^{\dagger} c_{i\alpha} 
+ \sum_{\langle \langle i,j \rangle\rangle}  \Delta_{ij} c_{i\alpha}^{\dagger} c_{j\alpha^{\prime}}
\notag
\\ & &
+ \dfrac{2 i}{3} \sum_{\langle i,j \rangle} \sum_{\alpha \alpha^{\prime}} c_{i\alpha}^{\dagger} c_{j\alpha^{\prime}} [ \lambda_{R} (\bm{s} \times \mathbf{\hat{d}}_{ij})_{z}]_{\alpha \alpha^{\prime}}.
\label{tb1}
\end{eqnarray} 
Here $\Delta_{ij}=i\lambda_{c_{i}} \nu_{ij} s_{z} /3 \sqrt{3}$, $c_{i\alpha}^{\dagger}$ creates an electron with spin polarization $\alpha$ at site $i$ that belongs to sublattice $A$ or $B$, and $\langle i,j \rangle$ $(\langle \langle i,j \rangle\rangle)$ runs over the nearest (second nearest) neighbouring sites. The second term is a staggered on-site potential, which takes into account the effective energy difference experienced by atoms at the lattice sites $A$ $(\eta_{c_{i}}=+1)$ and $B$ $(\eta_{c_{i}}=-1)$, respectively. The third and fourth terms represent the proximity-induced enhancement of the SOC due to a weak hybridization with the heavy atoms in WSe$_{2}$. 
The third 
term is the 
 valley-Zeeman SOC where $\nu_{ij}=+1$, if the second nearest hopping is anticlockwise with respect to the positive $z$ axis, and $\nu_{ij}=-1$ if it is clockwise. The  last term is the Rashba SOC parametrized by $\lambda_{R}$. It arises because the inversion symmetry is broken when the graphene sheet is placed on  top of WSe$_{2}$ as shown in Fig.\ref{f1} $(c)$. Also, $\mathbf{\hat{d}}_{ij} = \mathbf{d}_{ij} / \vert \mathbf{d}_{ij} \vert$, where $\bm{s}= (s_{x},s_{y},s_{z})$ is the Pauli spin matrix  and $\mathbf{d}_{ij}$ the vector connecting the sites $i$ and $j$ in the same sublattice. 

In Fig. \ref{f2} we plot the numerically evaluated energy dispersion of Eq. (\ref{tb1}) to better understand the characteristics of the induced intrinsic SOCs. Near the $K$ point, for $\lambda_{c_{i}}= \lambda_{R}= 0$, the band structure has linear band crossings near $k=0$ as can be seen from Fig. \ref{f2} (a). For 
$\lambda_{c_{i}} \neq 0$ and $\lambda_{R}=0$ 
the spectrum is gapless and the spin degeneracy is broken away from $k=0$, see Fig. \ref{f2} (b). Further, if only $\lambda_{R}$ is present, the spectrum is also gapless, cf. Fig. \ref{f2} (c). However, a gap is created when both  $\lambda_{c_{i}}$ and $\lambda_{R}$ are finite, cf. Fig. \ref{f2} (d).

We analyze the physics of electrons near the Fermi energy using a low-energy effective Hamiltonian derived from 
Eq. (\ref{tb1}) and a Dirac theory around the $K$ and $K^{\prime}$ valleys  \cite{rr23, r31, rr32}. It  reads
\begin{equation}
H^{s}_{\eta}=v_{F}(\eta\sigma_{x}p_{x}+\sigma_{y}p_{y})+\Delta\sigma_{z}+ \lambda \sigma_{0} s \eta +\lambda_{R}(\eta s_{y}\sigma_{x}-s_{x}\sigma_{y}). \label{e1}
\end{equation}
Here $\eta=\pm1$ denotes the valleys $K$ and $K^{\prime}$, $\Delta$ is the mass term that breaks the inversion symmetry, $\lambda =\lambda_{c_{i}}$ is the valley-Zeeman SOC strength, $\lambda_{R}$  the Rashba type SOC strength, ($\sigma_{x}$, $\sigma_{y}$, $\sigma_{z}$)  the Pauli matrix that corresponds to the pseudospin (i.e., $A-B$ sublattice), $\sigma_{0}$ is the unit matrix in the sublattice space,  and $v_{F}$ ($8.2 \times 10^{5}$ m/s) denotes the Fermi velocity of Dirac fermions. For simplicity, we neglect the intrinsic SOC $\lambda_{i}$ and consider  only the $\lambda_{R}> \lambda_{i}$ case. Also, we expect that small but finite values of $\lambda_{i}$ do not qualitatively affect our results as long as $\lambda >> \lambda_{i}$. Further, we will also neglect the $\Delta$ term in our numerical treatment because $ \lambda>> \Delta$.
\begin{figure}[t]
\centering
\includegraphics[width=8cm, height=8cm]{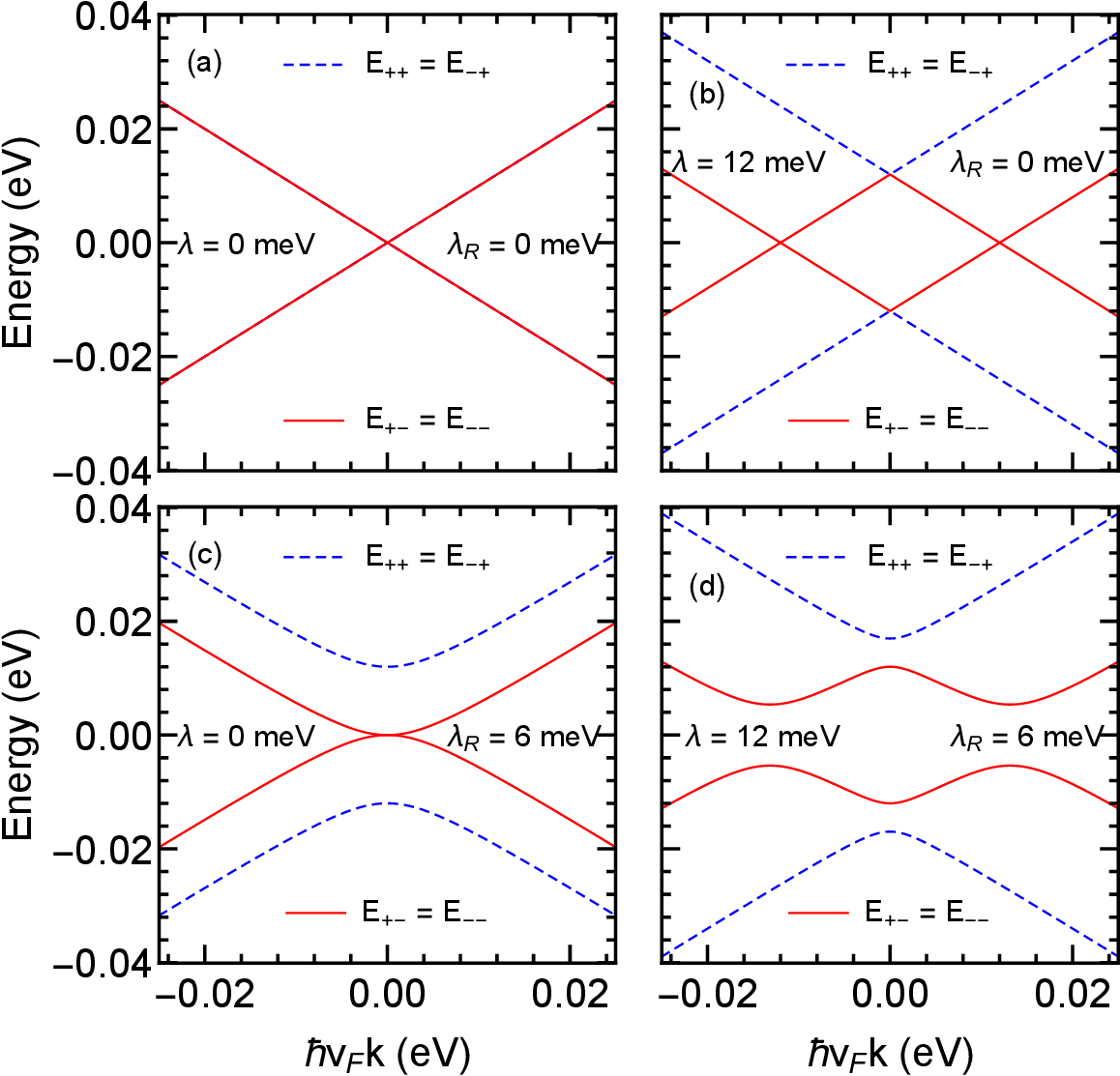}
\caption{Low-energy dispersion in a graphene/WSe$_{2}$ heterostructure for $\Delta=0$ and different combinations of 
$\lambda$ and $\lambda_{R}$.} 
\label{f3}
\end{figure}

Upon diagonalizing Eq. (\ref{e1}) we obtain the 
dispersion 
\begin{align}
\hspace*{-0.3cm}E_{\xi}(k)  &  = l 
\big[\Delta^{2}+\lambda^{2}+ \hslash^{2} v_{F}^{2} k^{2}+2\lambda_{R}^{2}+ 2 s \sqrt{\Upsilon}\,\big]^{1/2}, \label{e2}
\end{align}
where 
 $\Upsilon = \lambda_{R}^{2}\left(  \lambda_{R}^{2}-2\lambda\Delta\right)  + \hslash^{2} v_{F}^{2} k^{2} \left(\lambda_{R}^{2}+\lambda^{2}\right)  +\lambda^{2}\Delta^{2}$ and $\xi =\{l,s\}$. Further, $l= +1 (-1)$ denotes the conduction (valence) band and $s= +1 (-1)$ represents the spin-up (spin-down) branches. Notice that Eq. (\ref{e2}) has a valley degeneracy despite  the valley-Zeeman term.  The normalized eigenfunctions for both valleys are
\begin{equation}
\psi_{\xi}^{+} (k) = \dfrac{N_{\xi}^{+}}{\sqrt{S_{0}}}
\begin{pmatrix}
1\\
A_{\xi}^{\eta} e^{i\phi}\\
-i B_{\xi}^{\eta} e^{i\phi}\\
-i C_{\xi}^{\eta}  e^{2i\phi}
\end{pmatrix}
e^{i {\bf k} \cdot {\bf r}}\label{3},
\end{equation}
\begin{equation}
\psi_{\xi}^{-} (k) = \dfrac{N_{\xi}^{-}}{\sqrt{S_{0}}}
\begin{pmatrix}
- A_{\xi}^{\eta} e^{i\phi}\\
1\\
i C_{\xi}^{\eta} e^{2i\phi}\\
-i B_{\xi}^{\eta} e^{i\phi} 
\end{pmatrix}
e^{i {\bf k} \cdot {\bf r}}\label{4},
\end{equation}
respectively, with
\begin{align}
N_{\xi}^{\eta} &  = l\big[1 + ( A_{\xi}^{\eta}) ^{2} + ( B_{\xi}^{\eta}) ^{2} + ( C_{\xi}^{\eta}) ^{2}   \big ]^{-1/2},\label{e4}
\end{align}
 $S_{0}=L_{x}L_{y}$  the area of the sample, and $\phi = \tan^{-1}(k_{y}/k_{x})$. Further, 
$ A_{\xi}^{\eta} =   (E_{\xi}^{\eta} - \eta \Delta -\eta \lambda)   /  \hslash  v_{F} k$, $B_{\xi}^{\eta} = 2\lambda_{R}\big[ (E_{\xi}^{\eta})^{2} -(  \Delta + \lambda )  ^{2}\big]\big/  \hslash  v_{F} k \big[  (  E_{\xi}^{\eta} + \eta  \lambda ) ^{2}- \Delta^{2}- \hslash^{2}  v_{F}^{2} k^{2}\big]$, and $C_{\xi}^{\eta} =  2 \lambda_{R} (  E_{\xi}^{\eta} -\eta \Delta -\eta \lambda ) \big /  [( E_{\xi}^{\eta} + \eta \lambda )  ^{2}-\Delta^{2}-  \hslash^{2}  v_{F}^{2} k^{2}] $.
%
\begin{figure}[t]
\centering
\includegraphics[width=7.5cm, height=6cm]{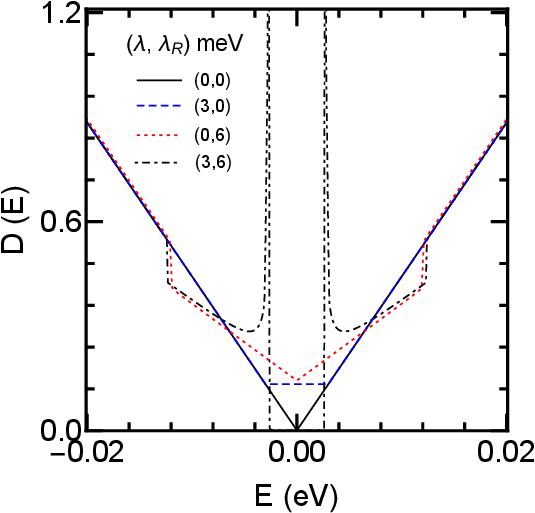}
\caption{Density of states for $(\lambda, \lambda_{R})= (0, 0)$, $(\lambda, \lambda_{R})= (3, 0)$ meV, $(\lambda, \lambda_{R})= (0, 6) $ meV and  $(\lambda, \lambda_{R})= (3, 6)$ meV. All cases are for $\Delta = 0$.}
\label{f4}
\end{figure}

We plot Eq. (\ref{e2})  in Fig. \ref{f3} for different combinations of the $\lambda$ and $\lambda_{R}$ terms whose realistic values  fall 
 in the ranges $5-6$ meV and $10-15$ meV, respectively, as determined experimentally in Ref. \onlinecite{nnn7}. Here, the larger values of SOCs are used just to see well-resolved bands splitting. For $\lambda= \lambda_{R}= 0$, the band structure has linear bands crossing near $k=0$ for both valleys as can be seen from panel (a). For $\lambda \neq 0$ and $ \lambda_{R}= 0$, the energy dispersion is spin non-degenerate and valley degenerate 
with a gapless behaviour 
as shown in panel (b). Further, the energy dispersion shows the gapless behaviour for $\lambda= 0$ and $ \lambda_{R} \neq 0$ whereas it is spin-split as seen from panel (c). However, for $\lambda$ and $\lambda_{R}$ finite, 
the Rashba coupling not only creates a  gap between the conduction and valence band, by mixing the spin-up and spin-down states, but also produces an avoided crossing, see Fig. \ref{f1} (d). The analytical form of the momentum $k_{1}$, at which an avoided crossing occurs, and of the gap $E_{g}=\Delta_1$ are 
\begin{equation}
k_{1}={1\over\hslash v_{F}}
\Big[\dfrac{(\lambda^{2}+\lambda \Delta)(\lambda^{2}+2 \lambda_{R}^{2}-\lambda \Delta)}{\lambda^{2}+ \lambda_{R}^{2}}\Big]^{1/2},
\end{equation}
\begin{equation}
\Delta_{1}=2\lambda_{R}\Big[\dfrac{\lambda^{2} +  \Delta(2 \lambda+\Delta)}{\lambda^{2}+ \lambda_{R}^{2}}\Big]^{1/2}.
\end{equation}

The density of states (DOS) per unit area corresponding to Eq. 
(\ref{e2}) is given by 
$D(E)=\sum_{\zeta} \delta(E-E_{\zeta})$ with $\left\vert  \zeta \right\rangle= \left\vert  \xi, \eta ,k \right\rangle$. For $\lambda_{R}=0$ it takes the simple form
\begin{equation}
D(E)=\dfrac{1}{2\pi \hslash^{2} v_{F}^{2}}\sum_{\xi} \left\vert \dfrac{E}{l}-s \lambda \right\vert \Theta(\dfrac{E}{l}-s \lambda-\Delta),
\end{equation}
and for $\Delta=\lambda=0$ 
the form
\begin{equation}
\hspace*{-0.3cm}D(E)=\dfrac{1}{2\pi \hslash^{2} v_{F}^{2}}\sum_{\xi} \left\vert \dfrac{E}{l}-s \lambda_{R} \right\vert \Theta(\dfrac{E}{l}-(s+1) \lambda_{R}).
\end{equation}

The DOS is shown in Fig. \ref{f4} for several values of $\lambda$ and $\lambda_{R}$. The black curve is for monolayer graphene, with  $\lambda=\lambda_{R}=0$, and is included for comparison. The $E_{+-}$ and $E_{++}$ dispersions give rise to a square root singularity at $E=\lambda \lambda_{R}/ \sqrt{\lambda^{2}+\lambda_{R}^{2}}$ and a step at $E= \sqrt{\lambda^{2}+ 4 \lambda_{R}^{2}}$, respectively, as shown by the black dot-dashed curve of Fig. \ref{f4}. The origin of the singularity is 
the mexican-hat  energy dispersion, cf. Fig. \ref{f3}. In addition, the step emerges from the bottom of the $E_{++}$ band and is a van Hove singularity associated with the dispersion flattening at this point. The square root singularity is calculated near the mexican-hat minimum $E=\lambda \lambda_{R}/ \sqrt{\lambda^{2}+\lambda_{R}^{2}}$ at which $D(E)$ reads 
\begin{equation}
D(E)= \dfrac{k_{1}}{4 \pi \hslash} \sqrt{\dfrac{2m^{\ast}}{E-\Delta_{1}}},
\end{equation}
\noindent with  $m^{\ast}= \lambda_{R} (\lambda^{2}+\lambda_{R}^{2})^{3/2}/ 2 v_{F}^{2}\lambda (\lambda^{2}+2 \lambda_{R}^{2})$ the effective mass and  $E_{+,-}=\Delta_{1}+(\hslash^{2}/2m^{\ast})(k-k_{1})^{2}$ the energy. This 
singularity is similar to that of the 
 one-dimensional density of states. In the limit $\lambda_{R}=0$ and $\lambda \neq 0$, the DOS has a finite value $\lambda/2 \pi \hslash^{2} v_{F}^{2}$ at $E=0$ (see blue dashed curve). For $E \geqslant \lambda$, it increases linearly with $E$. Also, for $\lambda=0$ and $\lambda_{R} \neq 0$, it is finite at $E=0$ but has a step at $E=2 \lambda_{R}$, see the red dotted curve.
\begin{figure}[t]
\centering
\includegraphics[width=7.5cm, height=6cm]{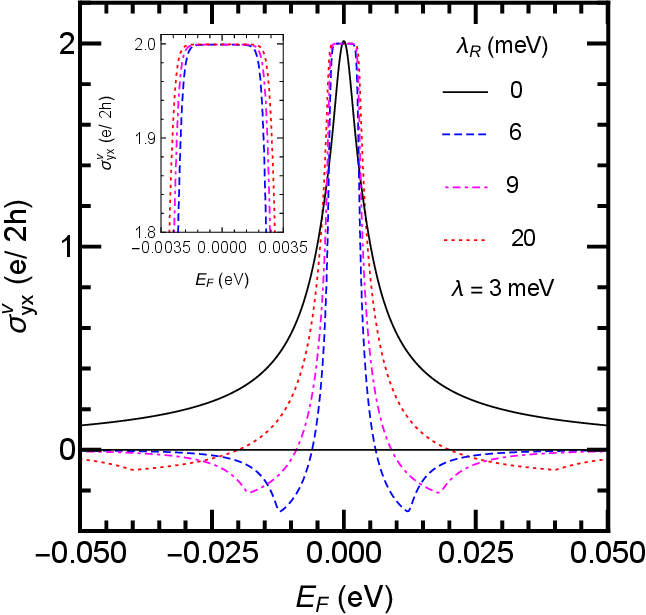}
\caption{Valley-Hall conductivity versus Fermi energy at $T =0.5$ K. For further clarity, the range -$0.35$ meV $\leq E_F\leq 0.35$ meV is shown in the inset without the $\lambda_R=0$ curve.}
\label{f5}
\end{figure}
 \begin{table*}
\caption{\label{tab1}Band gap energies involved in optical transitions, cf. Fig. \ref{f6}, for $\lambda=8$ meV, $\lambda_{R}=6$ meV, and  two 
values of $E_{F}$.}
\begin{ruledtabular}
\begin{tabular}{cccccccc}
Transition energies & Formula & \makecell{
$E_{F}=6.6$ meV\quad\quad $E_{F}=9.6$ meV} 
\\
\hline
\ \\
$\Delta_{1}$& $2 \lambda \lambda_{R}/\sqrt{\lambda^{2} + \lambda_{R}^{2}}$  
& \hspace*{-2cm}9.6 & \hspace*{-2cm}9.6 \\
\ \\
$\Delta_{2}$& $2 \sqrt{(4\lambda^{4} + 4 \lambda_{R}^{4}+9 \lambda^{2} \lambda_{R}^{2})/(\lambda^{2} + \lambda_{R}^{2})}$ & \hspace*{-2cm}41.2 & \hspace*{-2cm}41.2  \\
\ \\
$\Delta_{01}$& $2 \lambda$ & \hspace*{-2cm}16 & \hspace*{-2cm}16  \\
\ \\
$\Delta_{02}$& $2 \sqrt{\lambda^{2} +4 \lambda_{R}^{2}}$ & \hspace*{-2cm}28.8 & \hspace*{-2cm}28.8 \\
\ \\
$\Delta_{a}$& $2 \sqrt{2 \lambda^{2}+2\lambda_{R}^{2}+E_{F}^{2}-2M-2L}$ & \hspace*{-2cm}32.2 &  \\
\ \\
$\Delta_{b}$& $2 \sqrt{2 \lambda^{2}+2\lambda_{R}^{2}+E_{F}^{2}+2M+2L}$  & \hspace*{-2cm}50 & \hspace*{-2cm}57.4 \\
\ \\
&  \hspace*{-0.5cm}$M= \sqrt{(\lambda^{2} + \lambda_{R}^{2})E_{F}^{2}-\lambda^{2}\lambda_{R}^{2}}$, \quad  $L=\sqrt{\lambda_{R}^{4}+ (\lambda^{2} + \lambda_{R}^{2})(E_{F}^{2}+\lambda^{2}\pm 2M)}$
\end{tabular}
\end{ruledtabular}
\end{table*}

\section{ Conductivities} 

We consider a many-body system described by the Hamiltonian $H = H_{0} + H_{I} - \mathbf{R \cdot F}(t)$, where $H_{0}$ is the unperturbed part, $H_{I}$ is a binary-type interaction (e.g., between electrons and impurities or phonons), and $ \mathbf{- R \cdot F}(t)$ is the interaction of the system with the external field F(t) \cite{r34}. For conductivity problems we have $\mathbf{F}(t) = e \mathbf{E}(t)$, where $\mathbf{E}(t)$ is the electric field, $e$ the electron charge, $\mathbf{R} = \sum_i {\bf r}_{i}$ , and $\mathbf{r}_{i}$  the position operator of electron $i$. In the representation in which $H_{0}$ is diagonal the many-body density operator $\rho = \rho^{d} + \rho^{nd}$ has a diagonal part $\rho^{d}$ and a nondiagonal part $\rho^{nd}$. For weak electric fields and weak scattering potentials, for which the first Born approximation applies, the conductivity tensor has a diagonal part $\sigma_{\mu\nu}^{d}$ and a nondiagonal part $\sigma_{\mu\nu}^{nd}$; the total conductivity is $\sigma_{\mu\nu}^T = \sigma_{\mu\nu}^{d} + \sigma_{\mu\nu}^{nd}, \mu,\nu = x,y$. 

 In general we have two kinds of currents, diffusive and hopping, with $\sigma_{\mu\nu}^{d} = \sigma_{\mu\nu}^{dif} + \sigma_{\mu\nu}^{col}$, but usually only one of them is present. If  no magnetic field is present,  the hop-\\ping  term $\sigma_{\mu\nu}^{col}$ vanishes identically  and  only the term $ \sigma_{\mu\nu}^{dif} $ survives.  For  elastic scattering it is given  by \cite{r34} 
\begin{equation}
\sigma_{\mu\nu}^{d} (\omega) = \dfrac{\beta e^{2}}{S_{0}} \sum_{\zeta} f_{\zeta} (1 - f_{\zeta} ) \dfrac{v_{\nu\zeta}\, v_{\mu\zeta}\, \tau_{\zeta}}{1 + i\omega \tau_{\zeta}} , \label{c1}
\end{equation}
with $\tau_{\zeta}$  the momentum relaxation time, $\omega$ the frequency, 
and $v_{\mu\zeta}$ the diagonal matrix elements of the velocity operator. Further, $f_{\zeta} = \big[1 + \exp [\beta (E_{\zeta} - E_{F})]\big]^{-1}$ is the Fermi-Dirac distribution function, $\beta = 1/k_{B}T$ and $T$ the temperature.
\begin{figure}[t]
\centering
\includegraphics[width=7.5cm, height=6cm]{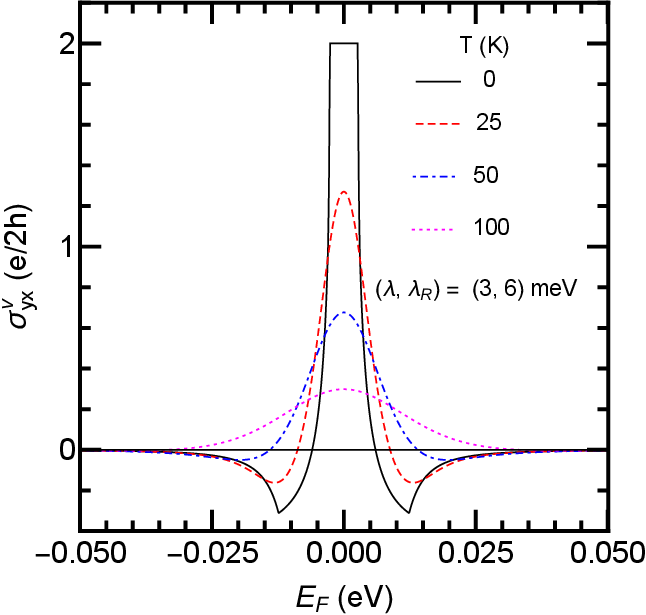}
\caption{Same as in Fig. \ref{f5} but for different values of $T$.}
\label{t01}
\end{figure}

Regarding the contribution $\sigma_{\mu\nu}^{nd}$ one can use the identity $f_{\zeta} (1 - f_{\zeta^{\prime}})\big[1 - \exp [\beta (E_{\zeta} - E_{\zeta^{\prime}})]\big] = f_{\zeta} - f_{\zeta^{\prime}}$ and cast the original form in the more familiar one \cite{r34}
%
\begin{equation}
\hspace*{-0.4cm}\sigma_{\mu\nu}^{nd} (\omega) =\dfrac{ i\hslash e^{2}}{S_{0}}\sum_{\zeta \neq \zeta^{\prime}} \dfrac{(f_{\zeta} - f_{\zeta^{\prime}}) \,v_{\nu\zeta\zeta^{\prime}} \,v_{\mu \zeta\zeta^{\prime}}}{(E_{\zeta} - E_{\zeta^{\prime}})(E_{\zeta} - E_{\zeta^{\prime}} + \hslash \omega - i \Gamma )} ,\label{c2}
\end{equation}
where the sum runs over all quantum numbers $ \zeta $ and $ \zeta^{\prime} $ with $\zeta \neq \zeta^{\prime}$. The infinitesimal quantity $\epsilon$ in the original form has been replaced by $\Gamma_{\zeta}$ to account for the broadening of the energy levels. In Eq. (\ref{c2}) $v_{\nu \zeta \zeta^{\prime}}$ and $v_{\mu \zeta \zeta^{\prime}}$ are the off-diagonal matrix elements of the velocity operator. The relevant velocity operators are given by $v_{x}=  \partial H / \hslash \partial k_{x}$ and $v_{y}=  \partial H / \hslash \partial k_{y}$. With $\zeta=\{l,s.k, \eta\}=\{\xi, k, \eta\}$ for brevity, they read 
\begin{equation}
\left \langle \zeta  \right\vert  v_{x} \left \vert  \zeta^{\prime}\right\rangle  =  v_{F} N_{\xi}^{\eta}N_{\xi^{\prime}}^{\eta} (D_{\xi,\xi^{\prime}}^{\eta} e^{i\phi} +F_{\xi,\xi^{\prime}}^{\eta} e^{-i\phi} ) \delta_{k,k^{\prime}}, \label{v1}
\end{equation}
\begin{equation}
\left \langle \zeta^{\prime}  \right\vert  v_{y} \left \vert  \zeta \right\rangle  = i  v_{F} N_{\xi}^{\eta}N_{\xi^{\prime}}^{\eta} ( D_{\xi,\xi^{\prime}}^{\eta} e^{-i\phi}  - F_{\xi,\xi^{\prime}}^{\eta} e^{i\phi}  ) \delta_{k,k^{\prime}}, \label{v2}
\end{equation}
where $D_{\xi,\xi^{\prime}}^{\eta}= A_{\xi^{\prime}}^{\eta}+ B_{\xi}^{\eta}  C_{\xi^{\prime}}^{\eta}$ and  $F_{\xi,\xi^{\prime}}^{\eta}= A_{\xi}^{\eta}+ B_{\xi^{\prime}}^{\eta}  C_{\xi}^{\eta}$.

We now calculate the conductivity $\sigma_{yx}^{nd}(i \omega) $  given by Eq. (\ref{c2}). Further, the velocity matrix elements (\ref{v1}) and (\ref{v2}) are diagonal in $k$, therefore $k$ will be suppressed in order to simplify the notation. The summation in Eq. (\ref{c2}) runs over all quantum numbers $\xi$,$\xi^{\prime}$, $\eta$, $\eta^{\prime}$, and $k$. The parameter $\Gamma_{\eta \eta^{\prime}}^{\xi \xi^{\prime}}$, that takes into account the level broadening, is assumed to be independent of the band and valley indices, i.e., $\Gamma_{\eta \eta^{\prime}}^{\xi \xi^{\prime}}=\Gamma$.  Using 
Eqs. (\ref{v1}) and (\ref{v2}) we can express Eq. (\ref{c2}) as
\begin{eqnarray}
\notag
\sigma_{yx}^{nd}(i \omega) & = & \dfrac{e^{2} \hslash^{2} v_{F}^{2}}{h} \sum_{ \xi \xi^{\prime}} \int dk k \,\dfrac{(N_{\xi}^{\eta}N_{\xi^{\prime}}^{\eta})^{2} (f_{\xi k}^{\eta}- f_{\xi^{\prime}k}^{\eta})}{ \Delta_{\xi \xi^{\prime}}^{\eta} \big[ ( \Delta_{\xi \xi^{\prime}}^{\eta}+\hslash \omega)^{2}+ \Gamma^{2}\big]}
\notag
\\ & &
\times  \big[\Delta_{\xi \xi^{\prime}}^{\eta}+\hslash \omega- i \Gamma) \big]\big[D_{\xi,\xi^{\prime}}^{\eta})^{2} - (F_{\xi,\xi^{\prime}}^{\eta})^{2}\big] \label{c5}
\end{eqnarray}
where $\Delta_{\xi \xi^{\prime}}^{\eta}= E_{\xi k}^{\eta} - E_{\xi^{\prime} k}^{\eta}$. Further, in the limit $\Gamma=\omega=0$, Eq. (\ref{c5}) reduces to
\begin{eqnarray}
\notag
\sigma_{yx}^{nd} & = & \dfrac{e^{2} \hslash^{2} v_{F}^{2}}{h} \sum_{ \xi \xi^{\prime}} \int dk k\, \dfrac{(N_{\xi}^{\eta}N_{\xi^{\prime}}^{\eta})^{2} (f_{\xi k}^{\eta}- f_{\xi^{\prime}k}^{\eta})}{(\Delta_{\xi \xi^{\prime}}^{\eta})^{2}}
\notag
\\ & &
\times \big[(D_{\xi,\xi^{\prime}}^{\eta})^{2} - (F_{\xi,\xi^{\prime}}^{\eta})^{2}\big] \label{c3}
\end{eqnarray}

In the valley-Hall effect 
 electrons from  regions near the inequivalent $K$ and $K^{\prime}$ valleys 
 flow to opposite transverse edges of the system, in the presence of SOCs  when  a longitudinal electric field is applied \cite{r37, nnn6}. The valley-Hall conductivity corresponding to Eq. (\ref{c5}) is defined by
\begin{equation}
\sigma_{yx}^{v} = \sum_{s s^{\prime}}\sigma_{yx}^{nd} (\eta=+,s, s^{\prime}) - \sigma_{yx}^{nd} (\eta=-,s, s^{\prime}).
\end{equation}
The  spin-Hall conductivity $\sigma_{yx}^{s}$, corresponding to Eq. (\ref{c5}), is finite only  when both the Kane-Mele and valley- Zeeman  SOCs are present. Hence, even in the presence of Rashba SOC, $\sigma_{yx}^{s}$ vanishes \cite{r30}. Since a spin current is defined by $ \bf{J}_{s} = ( \hslash / 2e) (\bf{J}_{\uparrow}-\bf{J}_{\downarrow})$, we have to multiply $\sigma_{yx}^{v}$ by $1/2e$ \cite{r5,r35}. Further, we find that charge Hall conductivity always vanishes
\begin{equation}
\sigma_{yx}^{c} = \sum_{\eta s s^{\prime}}\sigma_{yx}^{nd} (\eta,s, s^{\prime})=0
\end{equation}
%
\begin{figure}[t]
\centering
\includegraphics[width=7.5cm, height=6cm]{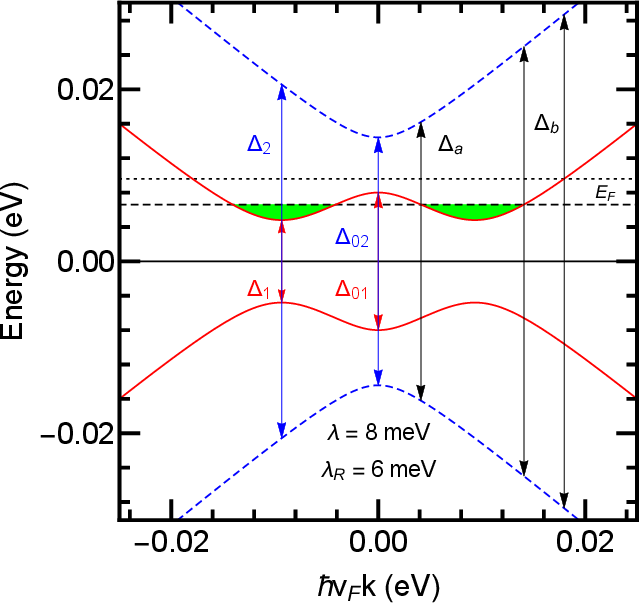}
\caption{Band structure near the Fermi energy $E_{F}$ in the presence of SOC terms for 
$\lambda=8$ meV and $\lambda_{R}=6$ meV. The black dashed and dotted lines show $E_{F}=6.6$ meV and $E_{F}=9.6$ meV. The various gap energies, indicated by $\Delta_1, \Delta_2$, etc.  are displayed in table \ref{tab1}. Notice that for $E_{F}=9.6$ meV the energy  $\Delta_a$ does not contribute to any transitions. }
\label{f6}
\end{figure}

The component $\sigma_{xx}^{nd}(i \omega)$ is also obtained from Eq. (\ref{c2}):
\begin{eqnarray}
\notag
\sigma_{xx}^{nd}(i \omega) & = & \dfrac{i e^{2} \hslash^{2} v_{F}^{2}}{ h} \sum_{\eta \xi \xi^{\prime}} \int dk k \,\dfrac{(N_{\xi}^{\eta}N_{\xi^{\prime}}^{\eta})^{2} (f_{\xi k}^{\eta}- f_{\xi^{\prime}k}^{\eta})}{ \Delta_{\xi \xi^{\prime}}^{\eta}\big[ ( \Delta_{\xi \xi^{\prime}}^{\eta}+\hslash \omega)^{2}+ \Gamma^{2}\big]}
\notag
\\ & &
\times \big[\Delta_{\xi \xi^{\prime}}^{\eta}+\hslash \omega- i \Gamma) \big]\big[D_{\xi,\xi^{\prime}}^{\eta})^{2} + (F_{\xi,\xi^{\prime}}^{\eta})^{2}\big]. \label{c6}
\end{eqnarray}

For $\lambda=0$ and $\lambda_{R}\neq 0$, Eq. (\ref{c5}) vanishes because the factor $(D_{\xi,\xi^{\prime}}^{\eta})^{2} - (F_{\xi,\xi^{\prime}}^{\eta})^{2}$ becomes zero, whereas Eq. (\ref{c6}) survives. Moreover, in the limit $\lambda=\lambda_{R}=0$, Eq. (\ref{c6}) reduces to the optical conductivity of pristine graphene, which is independent of $\hslash \omega$ and given by $e^{2}/2h$ \cite{r36}. 

We now consider  the diagonal component $\sigma_{xx}^{d}$  given by Eq. (\ref{c1}). 
Using Eq. (\ref{v1}),  with $\xi=\xi^{\prime}$, we obtain
\begin{eqnarray}
\notag
\sigma_{xx}^{d} (i \omega) & = & \dfrac{e^{2} v_{F}^{2} \beta}{\pi} \sum_{\eta  \xi } \int dk k \,(N_{\xi}^{\eta})^{4} f_{\xi k}^{\eta} (1- f_{\xi k}^{\eta}) 
\notag
\\ & &
\times \dfrac{(A_{ \xi}^{\eta}+ B_{ \xi}^{\eta}  C_{\xi}^{\eta})^{2}\, \tau_{\xi k}^{\eta}}{1+ i \omega \tau_{\xi k}^{\eta}} \label{c4}
\end{eqnarray}

At very low temperatures we can make the approximation $\beta f_{\xi k}^{\eta} (1- f_{\xi k}^{\eta})\approx\delta(E_{\xi}-E_{F})$ and  $\tau_{\xi k}^{\eta}=\tau_{\xi k_{F}}^{\eta}$ because all states untill the Fermi level are occupied.
\begin{figure}[t]
\centering
\includegraphics[width=9cm, height=9.5cm]{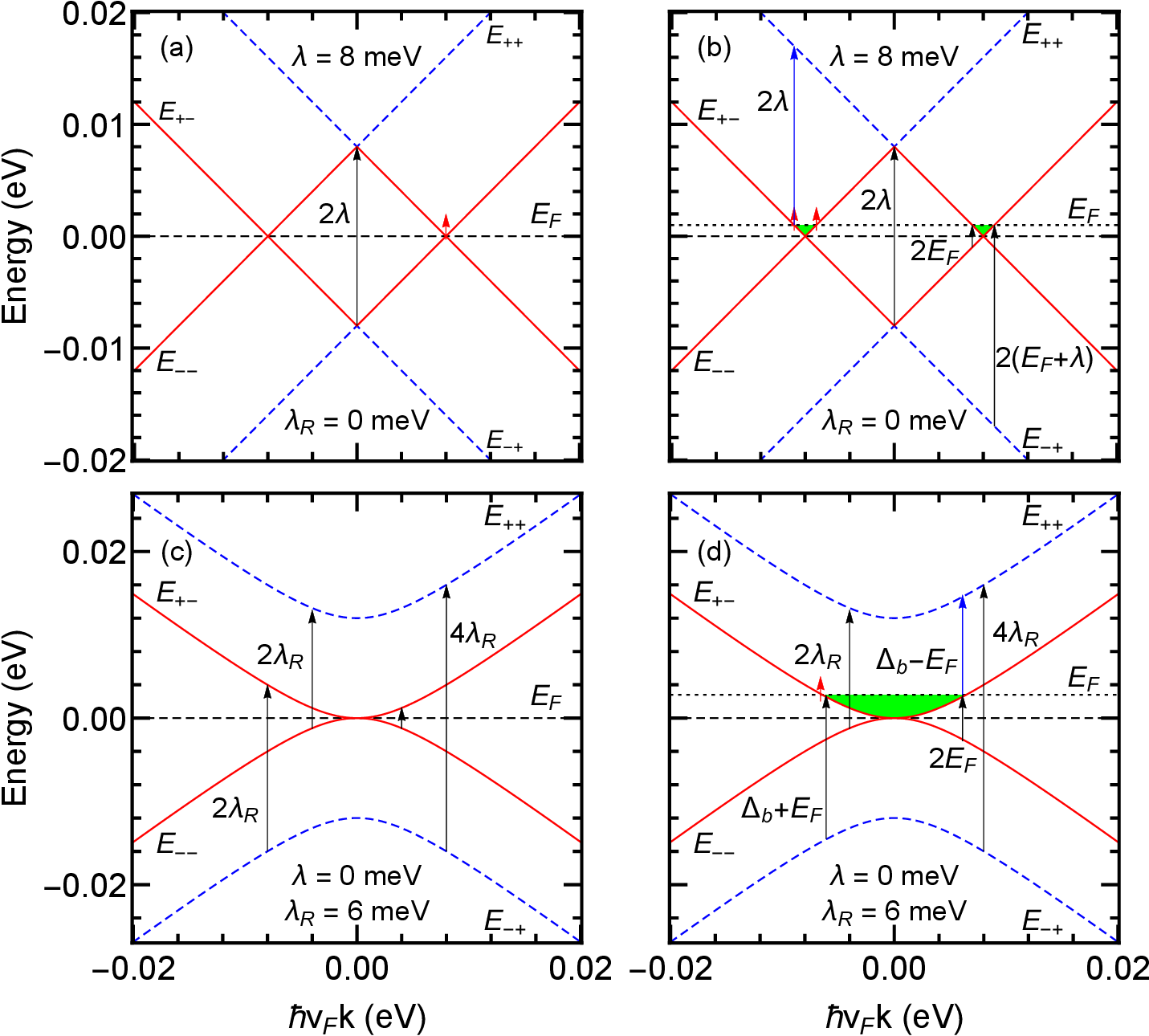}
\caption{Band structure near the Fermi energy $E_F$ in the presence of SOC terms for 
$\lambda=8$ meV and $\lambda_{R}=6$ meV. Black and red arrows represent  possible interband transitions. Red arrows indicate the Drude type intraband transitions. (a) $\lambda \neq 0$, $\lambda_{R},E_{F}= 0$. (b) $\lambda \neq 0$, $\lambda_{R}= 0$, $E_{F}=1$ meV. (c) $\lambda_{R} \neq 0$, $\lambda,E_{F}= 0$. (d) $\lambda_{R} \neq 0$, $\lambda_{R}= 0$, $E_{F}=2.8$ meV. }
\label{f7}
\end{figure}

In Fig. \ref{f5} we plot Eq. (\ref{c5}) in the dc limit ($\omega=0$) as a function of  $E_{F}$ for $\Gamma=0.2$ meV, $\lambda = 3$ meV and for different values of $\lambda_{R}$. When  $E_{F}$ is in the gap, i.e., in the range $-\lambda \lambda_{R}/\sqrt{\lambda^{2}+\lambda_{R}^{2}}$ $\leqslant E_{F} \leqslant  \lambda \lambda_{R}/\sqrt{\lambda^{2}+\lambda_{R}^{2}}$, the valley-Hall conductivity is quantized in units of $2e/2h$ similar to the case of gapped graphene and topological insulators \cite{r37, r38}. The reason is that the factor $\sum_{\eta \xi \xi^{\prime}} (N_{\xi}^{\eta}N_{\xi^{\prime}}^{\eta})^{2} [(D_{\xi,\xi^{\prime}}^{\eta})^{2} - (F_{\xi,\xi^{\prime}}^{\eta})^{2}]/(\Delta_{\xi \xi^{\prime}}^{\eta})^{2}$, called Berry curvature $\Omega(k)$, of Eq. (\ref{c5}) in the limit $\omega=0$ has a peak, which is well covered by occupied states for $E_{F}> \lambda \lambda_{R}/\sqrt{\lambda^{2}+\lambda_{R}^{2}}$. As a consequence, the valley-Hall conductivity approaches the quantized value.  For $\lambda \lambda_{R}/\sqrt{\lambda^{2}+\lambda_{R}^{2}}$  $\leqslant E_{F}\leqslant \lambda_{R}$, $\sigma_{yx}^{v}$ decreases with 
 $E_{F}$. Further, as can be seen, when  $E_{F}$ becomes comparable to $ \lambda_{R}$,                          
 a sign change occurs in the conductivity which later vanishes
 at higher values of $E_{F}$, $E_F>>\sqrt{\lambda^{2}+ 4 \lambda_{R}^{2}}$. The change in sign is due to the Rashba   coupling between the spin-up and spin-down bands. Furthermore, this off-diagonal term in  spin space permits 
 transitions between two conduction spin subbands (see Eq. (\ref{e2})), 
 that 
 could be interpreted as spin-flip transitions near the band touching. In addition, the coupling strength between opposite spin bands becomes weaker as $\lambda_{R}$ increases. As a result, the negative part of the conductivity due to the spin-up band diminishes and $\sigma_{yx}^{v}$ shows the usual behaviour of gapped graphene and topological insulators \cite{r37,r38}. Further, as can be seen in the inset, the band gap increases with $\lambda_{R}$. Also, the value of the conductivity at $E_F=0$ is due to  the finite one of $\Gamma$ (= 0.2 meV); if we take $\Gamma=0$, the conductivity diverges at $E_F=0$ but its overall qualitative behavior remains as shown.
\begin{figure}[t]
\centering
\includegraphics[width=7.5cm, height=10cm]{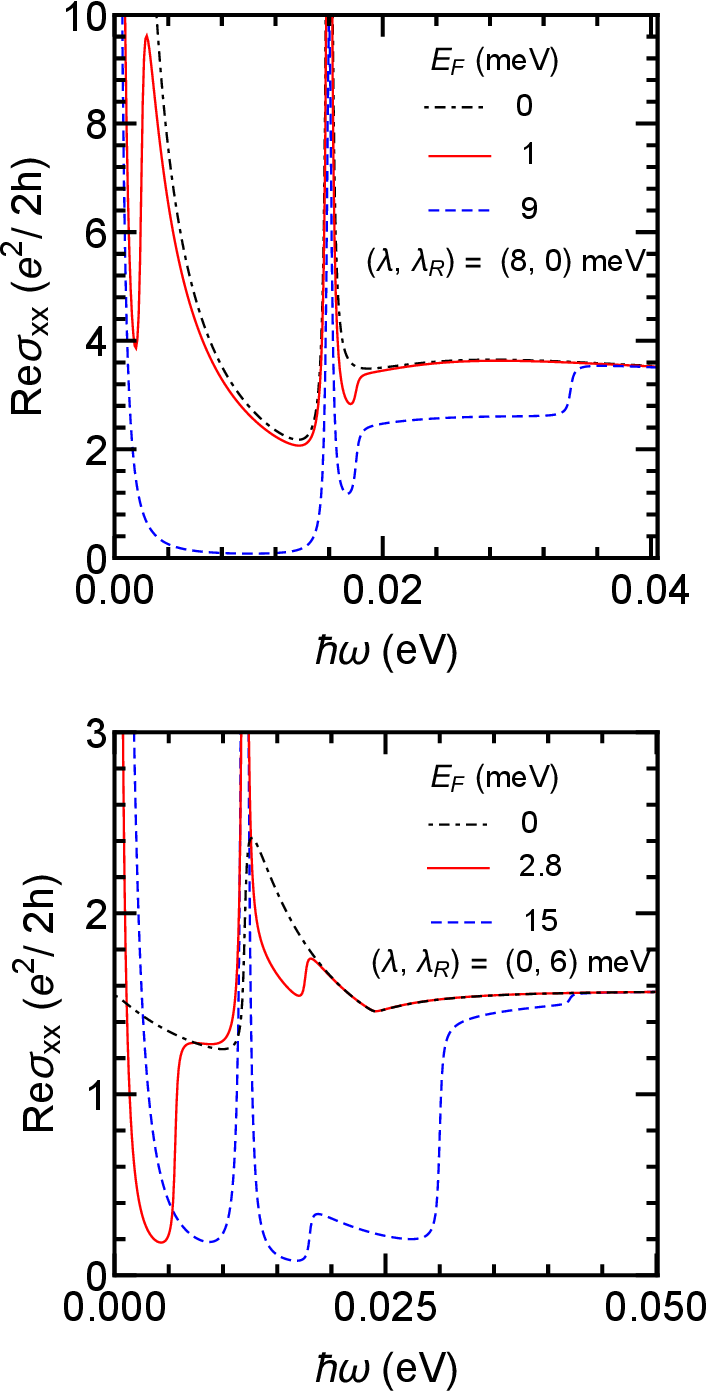}
\vspace{-0.2cm}
\caption{Real part of longitudinal conductivity versus photon energy at $T =0.5$ K. The upper panel is for $\lambda_{R}=0$ and the lower one is for $\lambda_{R} \neq 0$. }
\label{f8}
\end{figure}

We now take into account the effect of temperature $T$ on the valley-Hall conductivity contained in the Fermi function,  
which is independent of electron-phonon interaction in the first Born approximation \cite{r34}. 
The valley-Hall conductivity is evaluated numerically with the help of Eq. (\ref{c5}) and plotted in Fig. \ref{t01} for four values of $T$. We find a strong $T$ dependence, particularly when the Fermi level is in the gap. The quantization of the valley-Hall conductivity is destroyed at high values of $T$.  
  This occurs when  the thermal broadening $k_{B}T$  becomes comparable to the energy gap. Notice that the effect of temperature on $\sigma_{yx}^{v}$ is similar to that on  the spin-Hall conductivity in a graphene/MoS$_{2}$ heterostructure by considering  valley-Zeeman and Kane-Mele SOCs in the absence of the Rashba SOC.

Various transition 
energies, which play an important role in the optical conductivity, are shown in Fig. \ref{f6} for $\lambda,\lambda_{R} \neq 0$. Their analytical expressions are displayed in table \ref{tab1}.   Notice that for $E_{F}= 6.6$ meV,  the  energies $\Delta_{a}$ and $\Delta_{b}$, indicated with black arrows, become also important in optical transitions, since  $E_{F}$ crosses  the curve  $E_{+-}$ at two values of the momentum. 
However, for $E_{F}=9.6$ meV,  only $\Delta_{b}$ contributes to optical transitions because $E_{F}$ cuts $E_{+-}$ curve only at one value of the momentum. In Fig. \ref{f7}, we show possible allowed interband and intraband transitions by contrasting the  case $\lambda \neq 0, \lambda_{R}=0$ in the upper panels and the case $\lambda = 0, \lambda_{R} \neq 0$ in the lower panels. The blue arrows represent the interband transitions  $E_{+-}\rightarrow E_{++}$ for $0 < E_{F} < \lambda$ and $0 < E_{F} < \lambda_{R}$ as can be seen in Fig. \ref{f7} (b) and (d). The black arrows represent  the allowed interband transitions  $E_{-+} \rightarrow E_{+-} (E_{++})$ and $E_{--} \rightarrow E_{+-} (E_{++})$ for $E_{F} =0$ and $E_{F} \neq 0$, repectively, while the red arrows indicate 
intraband transitions that occur near $E_{F}$. 
\begin{figure}[t]
\centering
\includegraphics[width=8cm, height=5cm]{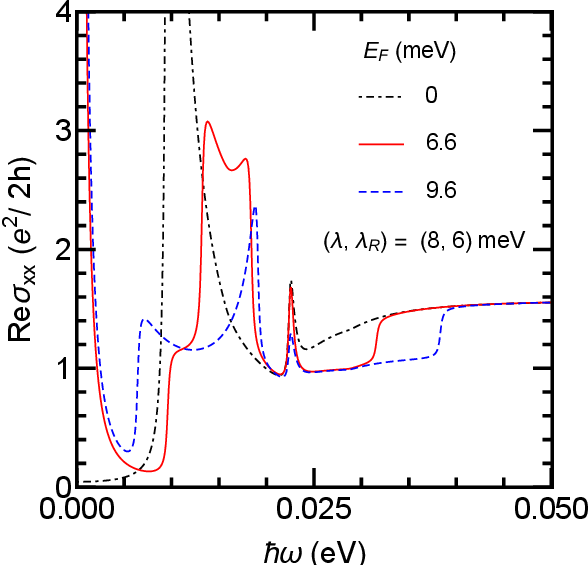}
\caption{Same as in Fig. \ref{f8} but for $\lambda, \lambda_{R} \neq 0$. }
\label{f9}
\end{figure}

Now we present results for the real part of Eqs. (\ref{c6}) and (\ref{c4}) $(\textrm{Re} \sigma_{xx}=\textrm{Re}\sigma_{xx}^{d}+\textrm{Re}\sigma_{xx}^{nd})$,  evaluated numerically, versus $\hslash \omega$ using a Lorentzian form of Dirac delta function and taking $\Gamma=0.2$ meV for $T \neq 0$. We start from the upper panel of Fig. \ref{f8} by considering the case $\lambda \neq 0$ and $\lambda_{R}=0$. The transitions are vertical for photon's momentum $q\sim 0$  and connect the filled valence band to empty conduction band, see Fig. \ref{f7} (a). For the case of $E_{F}=0$, intraband response appears due to the transition $E_{+-} \rightarrow E_{+-}$ and has a $\delta$ function form, centred around $\hslash \omega=0$, which broadens the peak when any kind of scattering is taken into account. Further, intraband 
responses occur when the Fermi level is located away from the Dirac point.  For $\hslash \omega=2 \lambda$ we obtain another Dirac delta peak due to the transition from $E_{--} \rightarrow E_{+-}$, which is also broadened through 
$i \pi \delta(x)=\lim_{\Gamma \to\ 0}(1/x - i \Gamma) $, cf. Eq. (\ref{c6}). For $ 0< E_{F}< \lambda$, the new absorption peaks appear at $\hslash \omega= 2E_{F}$ and $\hslash \omega= 2 (E_{F}+\lambda)$ due to the possible transitions   $E_{--}\rightarrow E_{+-}$ and $E_{-+}\rightarrow E_{+-}$. For $E_{F}>\lambda$, the absorption peaks disappear below $\hslash \omega < 2\lambda$ because the transition  $E_{--} \rightarrow E_{+-}$ is no longer possible due to the filling of states below the Fermi level that  are Pauli blocked. Further, the Drude peak persists at low $\hslash \omega$, but now two other pieces of interband transitions emerge with onsets at $\Delta_{a}+E_{F}$ and $\Delta_{b}+E_{F}$.
\begin{figure}[t]
\centering
\includegraphics[width=8cm, height=5cm]{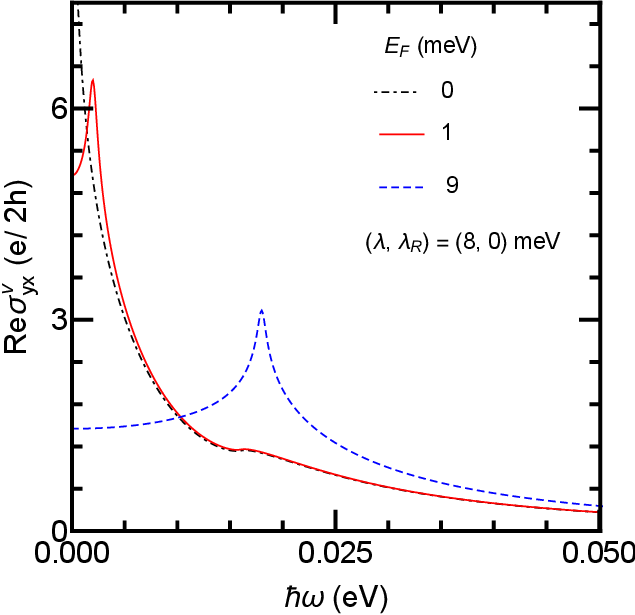}
\caption{Valley-Hall conductivity versus photon energy for $\lambda_{R}=0$ at $T =0.5$ K. }
\label{f10}
\end{figure}

In the lower panel of Fig. \ref{f8} we show the results for real part of the longitudinal conductivity for $\lambda=0$, $\lambda_{R} \neq 0$ for different values of $E_{F}$. For $E_{F}=0$, we can see that there is a peak at $2\lambda_{R}$ which is the separation between $E_{--}$ and $E_{+-}$ bands. In addition, there is a kink at $4\lambda_{R}$ due to the transition   $E_{-+} \rightarrow E_{++}$. As we increase the Fermi level, say, $0 <E_{F}< \lambda_{R}$ and $ E_{F}> \lambda_{R}$, the peak becomes sharper and we see a onset of a Drude contribution at low $\hslash \omega$ due to intraband transitions $E_{+-} \rightarrow E_{+-}$ and $E_{++} \rightarrow E_{++}$ in contrast to $E_{F}=0$ case (black dot-dashed curve). Further, for finite values of $E_{F}$, we see the steps at $2E_{F}$ similar to monolayer graphene ($\lambda=\lambda_{R}=0$) as well as features at $\Delta_{a}+E_{F}$, $\Delta_{b}-E_{F}$, and $\Delta_{b}+E_{F}$ above which we attain the flat absorption like pristine graphene \cite{r36}. Note that our results are similar to bilayer graphene \cite{r39,r40}. But here, the Rashba SOC, which allows the interband transitions between  opposite spin bands, gives rise to the absorption peaks, while these peaks in bilayer graphene are due to interlayer hopping between two graphene sheets.
\begin{figure}[t]
\centering
\includegraphics[width=8cm, height=9cm]{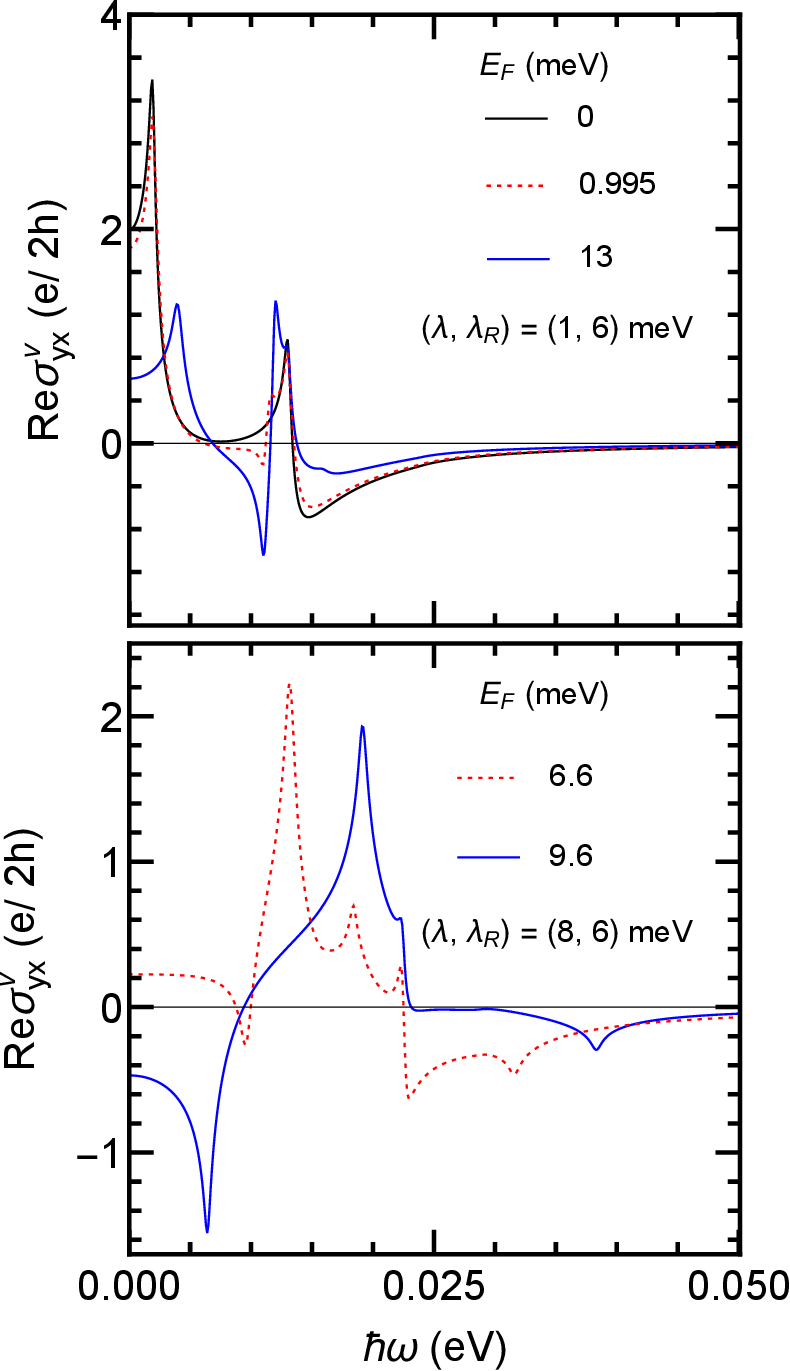}
\caption{Same as in Fig. \ref{f10} but with the upper panel  for $\lambda_{R} > \lambda$ and the lower one  for $\lambda_{R} < \lambda$. }
\label{f11}
\end{figure}

The real part of the longitudinal conductivity as a function of the photon energy, for $\lambda,\lambda_{R} \neq 0$, is show in Fig. \ref{f9} 
for several values of $E_{F}$: (i) just below the maximum of the mexican hat i.e. $ \lambda \lambda_{R}/(\lambda^{2} + \lambda_{R}^{2})^{1/2} < E_{F}< \lambda$ (ii) just above the mexican hat, i.e., for $ \lambda < E_{F}<  (\lambda^{2} + 2 \lambda_{R}^{2})^{1/2}$. For $E_{F}=0$ we find a large absorption peak at approximately $2\lambda_{R}$, which corresponds to transitions between the two square-root singularities of the DOS, see Fig. \ref{f4}, or transitions between the two minima of the mexican hat structures of the $E_{--}$ and $E_{-+}$ bands. As $E_{F}$ moves into the mexican hat, this feature  disappears because states below $E_{F}$ are occupied and, therefore,   Pauli blocked. Further, the major peaks are due to the transitions   $E_{+-}\rightarrow E_{++}$, $E_{--}\rightarrow E_{-+}$, $E_{--}\rightarrow E_{++}$ and $E_{-+}\rightarrow E_{++}$, respectively. The gap energies which contribute to the onset of these transition peaks are indicated in Fig. \ref{f6} and given analytically in table \ref{tab1}. Also, the conductivity retains the flat absorption at sufficiently higher values of $\hslash \omega$ similar to pristine graphene \cite{r36}.

Plots of the real part of $\sigma_{yx}^{v}$ for $E_{F}=0$ ( black dotdashed curve) and $E_{F}\neq 0$ (red and blue dashed curves) in the absence of Rashba SOC ($\lambda_{R}=0$) are shown in Fig. \ref{f10}. In the dc limit, the expected value of the valley-Hall conductivity is obtained as can be seen in Fig. \ref{f5} (black curve). If the system is illuminated by photons of frequency $\omega$, the amplitude of the absorption peaks is suppressed for $E_{F}=0$, while an increase in it is observed for $E_{F}\neq 0$. For $\hslash \omega= 2 \vert \lambda \vert $ a strong valley-Hall response is observed for $E_{F}\neq 0$. Therefore, it can be expected that a stronger valley-Hall response may be accessible when the photon energy is tuned to the valley-Zeeman SOC. For $\hslash \omega > 2 \vert \lambda \vert $, $\sigma_{yx}^{v}$   decreases rapidly and approaches   zero at sufficiently higher values of $\hslash \omega$.
\begin{figure}[t]
\centering
\includegraphics[width=8cm, height=6cm]{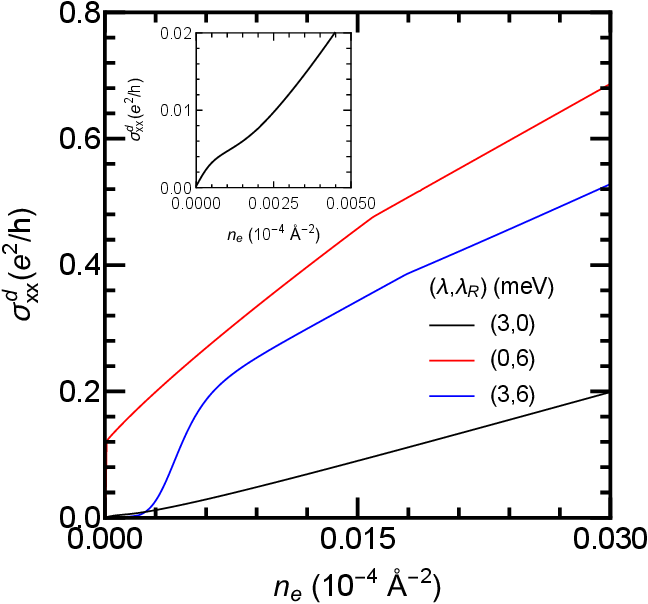}
\caption{ Longitudinal conductivity $\sigma^{d}_{xx}$ in units of $e^{2}/ h$ versus electron concentration $n_{e}$ for  different values of $\lambda$ and $\lambda_{R}$. For further clarity, the range $0-5$ meV is shown in inset. }
\label{f12}
\end{figure}

The real part of the valley-Hall conductivity is shown in Fig. \ref{f11} for several values of $E_{F}$. In the dc limit $(\omega=0)$, we obtain the quantized value of the valley-Hall conductivity (Re$ \sigma_{yx}^{v}=e/h $) for $E_{F}=0$ (black curve in the upper panel). If the system is subjected to photon of frequency $\omega$, an increase in the magnitude of the valley-Hall response is observed. The absorption peaks occur at the same onset energies as indicated in Fig. \ref{f6}. For example, the first peak appeared when $\hslash \omega= 2\Delta_{1}$ or transition between the minima of the $E_{--}$ and $E_{+-}$ bands. Further, the change in sign of the conductivity is due to the Rashba SOC, which is responsible for the coupling between spin-up and spin-down bands e.g., the transition from the maximum of mexican hat of $E_{--}$ band to the minimum of $E_{++}$ band around $k=0$. Furthermore, for finite values of  $E_{F}$ we obtain  new features in the optical spectrum due to the emergence of new transitions such as $E_{+-} \rightarrow E_{++}$, e.g.,  some features are completely removed due to Pauli blocking. Also, the valley-Hall response is diminished at sufficiently high frequencies. However, in the case of $\lambda_{R}<\lambda$ (lower panel), the difference among the optical transition energies is significantly enhanced due to larger values of $\lambda$ and new features emerge at the momenta at which $E_{F}$ crosses the $E_{+-}$ band (see Fig. \ref{f6}). Moreover, some of the optical transitions are no longer possible, e.g., $E_{--}\rightarrow E_{+-}$ when $E_{F}$ is just above the mexican hat because the  states below it are occupied and, therefore, Pauli blocked (blue curve).

\begin{figure}[t]
\centering
\includegraphics[width=7.5cm, height=5cm]{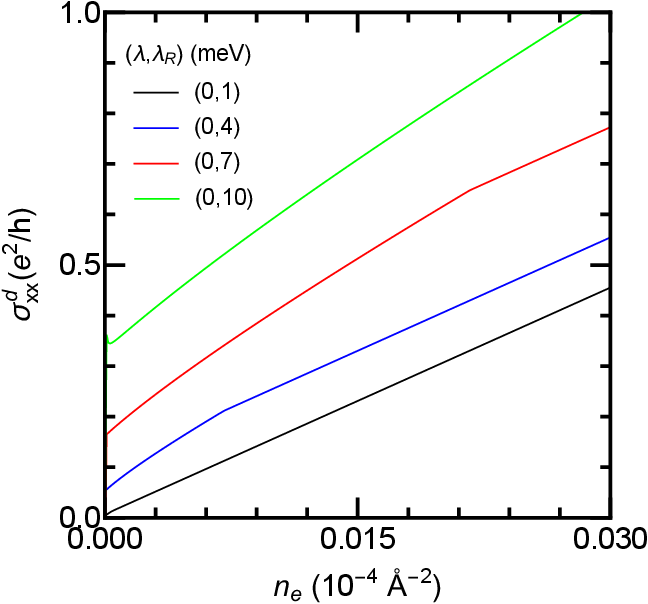}
\caption{ Same as in Fig. \ref{f12} but for different values of $\lambda_{R}$.}
\label{f13}
\end{figure}

In Fig. \ref{f12} we plot $\sigma_{xx}^d$, from  Eq. (\ref{c4}), by evaluating it numerically versus electron concentration ($n_{e}$) and using the expression of $\tau$  given in Appendix A but evaluated at the Fermi level, $k=k_F$. The conductivity  increases with  $E_{F}$ and therefore with the carrier density $n_e$. The diffusive conductivity increases  linearly with $n_e$ but cusp-like features appear when $E_{++}$ band begin to occupied at specific values of $n_{e}$ in contrast to pristine graphene \cite{r41,r42}. This behaviour makes graphene/WS$_{2}$ a suitable candidate for charge switches contrary to pristine graphene. The screening effect becomes significantly weaker when only the $\lambda$ term is present. Moreover, the conductivity shown in Fig. \ref{f13} increases in the low-density regime for $\lambda=0$ and $\lambda_{R} \neq 0$  as compared to  the $\lambda \neq 0$, $\lambda_{R}= 0$ and  $\lambda, \lambda_{R} \neq 0$ case. In the limit $\lambda=\lambda_{R}=0$ we obtain the result similar to pristine graphene \cite{r41,r42}.

\section{summary and conclusion}

We studied the energy dispersion of graphene/WSe$_{2}$ heterostructures by using a TB model in the presence of valley-Zeeman and Rashba  SOCs. 
 We found that the effective Hamiltonian (2) derived from the TB one (1) nicely captures the low-energy physics near the $K$ and $K^{\prime}$ valleys. We demonstrated that the density of states has a  finite value around $E=0$ in both cases $\lambda \neq 0, \lambda_{R}=0$ and $\lambda = 0, \lambda_{R} \neq 0$. In addition, it has a square root singularity when both $\lambda$ and $\lambda_{R}$ are present.  This singularity is similar to that in biased bilayer graphene; however, here it is due to the Rashba SOC  whereas in biased bilayer graphene it is due to interlayer hopping. We also found that the ac and dc valley-Hall conductivities change sign in the presence of the $\lambda_{R}$ term, which leads to interband transitions. Also, the band gap is enhanced by increasing the strength $\lambda_{R}$. Further, for $\lambda_{R} >> \lambda$ the valley-Hall conductivity exhibits a  behaviour similar to that in gapped graphene and topological insulators \cite{r37,r38}. The screening effect in the diffusive conductivity is dominant  only when the Rashba SOC is present, whereas it is significantly suppressed for $\lambda \neq 0,\lambda_{R}= 0$. Also, the conductivity increases with $\lambda_{R}$ in the low- and high-density regimes, see Fig. \ref{f13}. 

The dc valley-Hall conductivity changes sign when $E_F$ is comparable to $\lambda_R$ and vanishes at higher values of $E_F$, cf. Fig. 5. It also
exhibits a strong temperature dependence when the Fermi level in the gap, cf. Fig. 6.

The intraband response of the ac longitudinal conductivity for $\lambda_{R}=0$ (see upper panel of Fig. \ref{f8}) shifts towards lower photon energies when $E_F$ increases  compared to $\lambda_{R} \neq 0$ (see lower panel of Fig. \ref{f8} and Fig. \ref{f9}). We  also noted the switching on and off of the Drude response when the Fermi energy is varied (see Fig. \ref{f9}), which may be of interest in technological applications. In addition, for $\lambda,\lambda_{R} \neq 0$ new onsets in the  optical conductivity appear due to the shifting of the Fermi level through the mexican hat structure (see Figs. \ref{f9} and \ref{f11}), which may be a promising feature in  optical experiments. Our findings may be pertinent to developing future spintronics and valleytronics devices such as field-effect tunnelling transistors, 
memory devices,  phototransistors, etc.  

\acknowledgments

M. Z. and P. V. acknowledge the support of the Concordia University Grant No. VB0038 and a Concordia University Graduate Fellowship. The work of M. T. was supported by Colorado State University.

\appendix*

\begin{widetext}

\section{Relaxation time}

The relaxation time is generally a function of the incoming electron's wave vector and at low temperatures only states near the Fermi level will contribute to transport and single-particle properties. Below we  provide expressions for the relaxation time at the Fermi energy in the limiting cases $\Delta, \lambda \neq 0, \lambda_{R}=0$ and $\Delta, \lambda = 0, \lambda_{R} \neq 0$, because in these cases the summation over final states can be performed analytically.  Within the first Born approximation the standard formula for the momentum relaxation  time has the form
\begin{eqnarray}
\dfrac{1}{\tau_{\zeta}}=\dfrac{1}{\tau_{\xi k}^{\eta}} = \dfrac{2 \pi n_{i}}{\hslash} \sum_{\xi^{\prime}, \eta^{\prime}, k^{\prime}}  \vert \left\langle \xi, \eta, k  \right\vert  U(\mathbf{r}) \left\vert  \xi^\prime, \eta^{\prime}, k^{\prime} \right\rangle \vert^{2}  
 \delta(E_{\xi k}-E_{\xi^{\prime}k^{\prime}}) (1-\cos \theta), \label{t1}
\end{eqnarray}
where $U(\mathbf{r})$ is the impurity potential, $ n_{i}$  the impurity density, and $\theta$  the angle between the initial $k$ and final $k^{\prime}$ wave vectors. Equation~ (\ref{t1}) holds only for  elastic scattering $(\xi=\xi^{\prime},\eta=\eta^{\prime}, k=k^{\prime})$ and for central 
potentials $U(\mathbf{r})$ i.e. $U(\mathbf{r})=U(r)$. The results for two types of impurity potentials are as follows.

 {\it Short-range impurities.} We have $U(\mathbf{r})= U_{0} \delta(\mathbf{r}-\mathbf{r_{i}})$ where $\mathbf{r}$ and $\mathbf{r_{i}}$ are the position vectors of the electron and impurity, respectively, and $U_{0}$ is the strength of potential. In this case $U(\mathbf{q})= U_{0}$ is the Fourier transform of $U(\mathbf{r})= (1/\sqrt{L_{x}L_{y}})\sum_{q} U(q) e^{i \mathbf{q.r}}$ with $\vert \mathbf{q} \vert= 2k \sin(\theta/2)$. 
 The results are:

 i) $\lambda_{R}=0$
\begin{eqnarray}
\dfrac{1}{\tau_{s k_{F}}^{\eta}} = \dfrac{V_{0}^{2} n_{i}(N_{s}^{\eta})^{4}}{\hslash} \dfrac{\sqrt{\Delta^{2}+ (\hslash v_{F} k)^{2}}}{(\hslash v_{F})^{2}}\, \big[(A_{s}^{\eta})^{4}- (A_{s}^{\eta})^{2}+1\big].  \label{t2}
\end{eqnarray}
In the limit $\Delta, \lambda=0$, the above result reduces to graphene's scattering time Eq. (24) of Ref. \cite{r41}
\begin{equation}
\dfrac{1}{\tau_{ k_{F}}} =\dfrac{V_{0}^{2} n_{i}k}{4 \hslash^{2} v_{F}}.
\end{equation}
 Also, for $\lambda=0$ Eq. (\ref{t2}) agrees with the result for topological insulators \cite{r38}.

ii) $\Delta, \lambda=0$

\begin{eqnarray}
\dfrac{1}{\tau_{s k_{F}}^{\eta}} & = & \dfrac{V_{0}^{2} n_{i}(N_{s}^{\eta})^{4} \sqrt{\lambda_{R}^{2}+ (\hslash v_{F} k)^{2}} }{\hslash^{3} v_{F}^{2}} \,\Big [[(A_{s}^{\eta})^{2}+(B_{s}^{\eta}]^{2}]^{2}+ (C_{s}^{\eta})^{4}- [1+ (C_{s}^{\eta})^{2}] [(A_{s}^{\eta})^{2}+(B_{s}^{\eta})^{2}]
\\ & &
\notag
\hspace{4.5cm}-2 (C_{s}^{\eta})^{2} +1\Big].  \label{t3}
\end{eqnarray}

{\it  Long-range impurities.} We assume $U(\mathbf{r})= e Q e^{-k_{s}r}/4 \pi \epsilon_{0} \epsilon r$, where $k_{s}$ is the screening wave vector, $Q$ is the charge of the impurity, and $\epsilon$  the dielectric constant. In this case $U(q)= 2\pi U_{0}/\sqrt{k_{s}^{2}+q^{2}}$ with $U_{0}= eQ/4\pi \epsilon_{0}\epsilon$. The results are:

 i) $\lambda_{R}=0$
 
\begin{eqnarray}
\dfrac{1}{\tau_{s k_{F}}^{\eta}} & = & \dfrac{V_{0}^{2} n_{i}(N_{s}^{\eta})^{4} \sqrt{\Delta^{2}+ (\hslash v_{F} k)^{2}} }{2 \hslash^{3} v_{F}^{2} k^{2}}\, \Big[[1+(A_{s}^{\eta})^{4}]\big[1-\dfrac{a_{s}}{\sqrt{a_{s}^2+1}}\big]+ 2 (A_{s}^{\eta})^{2} \big[ 2 a_{s}^2-\dfrac{a_{s}(2 a_{s}^{2}+1)}{\sqrt{a_{s}^2+1}} \big]\Big].  \label{t4}
\end{eqnarray}

In the limit $\Delta=\lambda=0$ we set $a_{s}= k_{s}/2k$ 
and obtain the relaxation time in pristine graphene \cite{r43}
\begin{equation}
\dfrac{1}{\tau_{ k_{F}}}  = \dfrac{V_{0}^{2} n_{i}\, (a_{s}-\sqrt{a_{s}^2+1})^{2}}{4 \hslash^{2} v_{F}  k}. 
\end{equation}
Moreover, for $\lambda=0$ Eq. (\ref{t4}) gives the relaxation time for topological insulators \cite{r38}.

ii) $\Delta, \lambda=0$

\begin{eqnarray}
\dfrac{1}{\tau_{s k_{F}}^{\eta}} & = & \dfrac{V_{0}^{2} n_{i}(N_{s}^{\eta})^{4} \sqrt{\lambda_{R}^{2}+ (\hslash v_{F} k)^{2}} }{2 \hslash^{3} v_{F}^{2} k^{2}}\,\Big [1+[(A_{s}^{\eta})^{2}+(B_{s}^{\eta})^{2}]^{2} + (C_{s}^{\eta})^{4} [1-\dfrac{a_{s}}{\sqrt{a_{s}^2+1}}]
\\ &&
\notag
 + 2 [1+ (C_{s}^{\eta})^{2}] [A_{s}^{\eta})^{2}+(B_{s}^{\eta})^{2}] \big [2 a_{s}^2-\dfrac{a_{s}(2 a_{s}^{2}+1)}{\sqrt{a_{s}^2+1}} \big]
\\ &&
\notag
- 2 (C_{s}^{\eta})^{2} \big[ 1+\dfrac{a_{s}}{\sqrt{a_{s}^2+1}} + 8 a_{s}^{3} \sqrt{a_{s}^2+1} -8 a_{s}^{2}(2a_{s}^{2}+1)  \big]\Big ].  \label{t5}
\end{eqnarray}

\end{widetext}


\begin{thebibliography}{99}

\bibitem{r1} A. H. C. Neto, F. Guinea, N. M. R. Peres, K. S. Novoselov, and A. K. Geim, Rev. Mod. Phys. {\bf 81}, 109 (2009).

\bibitem{r2} N. Tombros, C. Jozsa, M. Popinciuc, H. T. Jonkman, and B. J. van Wees, Nature {\bf 448}, 571 (2007).

\bibitem{r3} J. Ingla-Ayn\'{e}s, M. H. D. Guimar\~{a}es, R. J. Meijerink, P. J. Zomer, and B. J. van Wees, Phys. Rev. B {\bf 92}, 201410(R) (2015).

\bibitem{r4} M. Dr\"{o}geler, C. Franzen, F. Volmer, T. Pohlmann, L. Banszerus, M. Wolter, K. Watanabe, T. Taniguchi, C. Stampfer, and B. Beschoten, Nano Lett. {\bf 16}, 3533 (2016).

\bibitem{r5} C. L. Kane and E. J. Mele, Phys. Rev. Lett. {\bf 95}, 226801 (2005).

\bibitem{r6} Z. Qiao, S. A. Yang, W. Feng, W.-K. Tse, J. Ding, Y. Yao, Y. Wang and
Q. Niu, Phys. Rev. B {\bf 82}, 161414 (2010); C.-X. Liu, S.-C. Zhang and X.-L. Qi, Annu. Rev. Condens. Matter Phys. {\bf 7}, 301 (2016); Y. Ren, Z. Qiao and Q. Niu, Rep. Prog. Phys. {\bf 79}, 066501 (2016); H. Weng, R. Yu, X. Hu, X. Dai and Z. Fang, Adv. Phys. {\bf 64}, 227 (2015).

\bibitem{r7} A. H. C Neto, and F. Guinea, Phys. Rev. Lett. {\bf 103}, 026804 (2009).

\bibitem{r8} C. Weeks, J. Hu, J. Alicea, M. Franz and R. Wu, Phys. Rev. X {\bf 1}, 021001 (2011).

\bibitem{r9} J. Ding, Z. Qiao, W. Feng, Y. Yao and Q. Niu, Phys. Rev. B {\bf 84}, 195444 (2011).

\bibitem{r10} J. Hu, J. Alicea, R. Wu and M. Franz, Phys.Rev. Lett. {\bf 109}, 266801 (2012).

\bibitem{r11} D. Ma, Z. Li and Z. Yang, Carbon {\bf 50}, 297 (2012).

\bibitem{r12} K.-H. Jin and S.-H. Jhi, Phys. Rev. B {\bf 87}, 075442 (2013).

\bibitem{r13} A. Ferreira, T. G. Rappoport, M. A. Cazalilla and A. H. C. Neto, Phys. Rev. Lett. {\bf 112}, 066601 (2014).

\bibitem{r14} A. A. Kaverzin and B. J. van Wees, Phys. Rev. B {\bf 91}, 165412 (2015).

\bibitem{r15} J. Balakrishnan, G. K. W. Koon, M. Jaiswal, A. H. C. Neto and B. C. Ozyilmaz, Nat. Phys. {\bf 9}, 284 (2013).

\bibitem{r16} X. Hong, S.-H. Cheng, C. Herding, and J. Zhu, Phys. Rev. B {\bf 83}, 085410 (2011).

\bibitem{r17} Z. Jia, B. Yan, J. Niu, Q. Han, R. Zhu, D. Yu and X. Wu, Phys. Rev. B {\bf 91}, 085411 (2015).

\bibitem{r18} U. Chandni, E. A. Henriksen and J. P. Eisenstein, Phys. Rev. B {\bf 91}, 245402 (2015).

\bibitem{rr19} Y.-C. Lin, N. Lu, N. Perea-Lopez, J. Li, Z. Lin, X. Peng, C. H. Lee, C. Sun, L. Calderin, P. N. Browning, M. S. Bresnehan, M. J. Kim, T. S. Mayer, M. Terrones, and J. A. Robinson, ACS Nano {\bf 8}, 3715 (2014).

\bibitem{rr20} M.-Y. Lin, C.-E. Chang, C.-H. Wang, C.-F. Su, C. Chen, S.-C. Lee, and S.-Y. Lin, Appl. Phys. Lett. {\bf 105}, 073501 (2014).

\bibitem{rr21} A. Azizi, S. Eichfeld, G. Geschwind, K. Zhang, B. Jiang, D. Mukherjee, L. Hossain,A. F. Piasecki, B. Kabius, J. A. Robinson, and N. Alem, ACS Nano {\bf 9}, 4882 (2015).

\bibitem{rr22} Y. Kim, D. Choi, W. J. Woo, J. B. Lee, G. H. Ryu, J. H. Lim, S. Lee, Z. Lee, S. Im, J.-H. Ahn, W.-H. Kim, J. Park, and H. Kim, Appl. Surf. Sci. {\bf 494}, 591 (2019).

\bibitem{rr23}  A. M. Alsharari, M. M. Asmar, and S. E. Ulloa, Phys. Rev. B {\bf 98}, 195129 (2018).

 \bibitem{rr24} L. A. Benitez, J. F. Sierra, W. S. Torres, A. Arrighi, F. Bonnell, M. V. Costache, and S. O. Valenzuera, Nat. Phys. {\bf 14}, 303 (2018).
 
 \bibitem{rr25} A. W. Cummings, J. H. Garcia, J. Fabian, and S. Roche, Phys. Rev. Lett. {\bf 119}, 206601 (2017).

\bibitem{r19} W. Han, R. K. Kawakami, M. Gmitra, and J. Fabian, Nat. Nanotechnol. {\bf 9}, 794 (2014).

\bibitem{r20} K. F. Mak, C. Lee, J. Hone, J. Shan, and T. F. Heinz, Phys. Rev. Lett. {\bf 105}, 136805 (2010).

\bibitem{r21} A. Korm\'{a}nyos, G. Burkard, M. Gmitra, J. Fabian, V. Z\'{o}lyomi, N. D. Drummond, and V. Fal\'{k}o, 2D Mater. {\bf 2}, 022001 (2015).

\bibitem{r22} C.-P. Lu, G. Li, K. Watanabe, T. Taniguchi, and E. Y. Andrei, Phys. Rev. Lett. {\bf 113}, 156804 (2014).

\bibitem{r23} S. Larentis, J. R. Tolsma, B. Fallahazad, D. C. Dillen, K. Kim, A. H. MacDonald, and E. Tutuc, Nano Lett. {\bf 14}, 2039 (2014).

\bibitem{nnn1} L. Banszerus,T. Sohier, A. Epping,F. Winkler, F. Libisch, F. Haupt, K. Watanabe, T. Taniguchi, K. Muller-Caspary, N. Marzari, F. Mauri, B. Beschoten, and C. Stampfer, arXiv: 1909.09523.

\bibitem{r24} S. Bertolazzi, D. Krasnozhon, and A. Kis, ACS Nano {\bf 7}, 3246 (2013).

\bibitem{r25} K. Roy, M. Padmanabhan, S. Goswami, T. P. Sai, G. Ramalingam, S. Raghavan, and A. Ghosh, Nat. Nanotechnol. {\bf 8}, 826 (2013).

\bibitem{r26} W. Zhang, C.-P. Chuu, J.-K. Huang, C.-H. Chen, M.-L. Tsai, Y.-H. Chang, C.-T. Liang, Y.-Z. Chen, Y.-L. Chueh, J.-H. He,  M.-Y. Chou, and L.-J. Lib, Sci. Rep. {\bf 4}, 3826 (2014).

\bibitem{r27} N. A. Kumar, M. A. Dar, R. Gul, and J. Baek, Mater. Today {\bf 18}, 286 (2015).

\bibitem{rr26} L. Britnel, R. V. Gorbachev, R. Jalil, B. D. Belle, F. Schedin, A. Mishchenko, T. Georgiou, M. I. Katsnelson, L. Eaves, S. V. Morozov, N. M. R. Peres, J. Leist, A. K. Geim1, K. S. Novoselov and L. A. Ponomarenko, Science {\bf 335}, 947 (2012).

\bibitem{rr27} A. Mishchenko, J. S. Tu, Y. Cao, R. V. Gorbachev, J. R. Wallbank, M. T. Greenaway, V. E. Morozov, S. V. Morozov, M. J. Zhu, S. L. Wong, F. Withers, C. R. Woods, Y-J. Kim, K. Watanabe, T. Taniguchi, E. E. Vdovin, O. Makarovsky, T. M. Fromhold, V. I. Fal’ko, A. K. Geim, L. Eaves and K. S. Novoselov, Nat. Nanotechnol. {\bf 9}, 808 (2014).

\bibitem{rr28} K. Roy, M. Padmanabhan, S. Goswami, T. P. Sai, G. Ramalingam, S. Raghavan and A. Ghosh, Nat. Nanotechnol. {\bf 8}, 826 (2013).

\bibitem{nnn2} A. David, P. Rakyta, A. Kormányos, and Guido Burkard, Phys. Rev. B {\bf 100}, 085412 (2019).

\bibitem{nnn3} Y. Li and M. Koshino, Phys. Rev. B {\bf 99}, 075438 (2019).

\bibitem{r28} A. Avsar, J. Y. Tan, T. Taychatanapat, J. Balakrishnan, G. K. W. Koon, Y. Yeo, J. Lahiri, A. Carvalho, A. S. Rodin, E. C. T. O\'{}Farrell,  G. Eda, A. H. C. Neto and B. \"{O}zyilmaz, Nat. Commun. {\bf 5}, 4875 (2014).

\bibitem{r29} M. Gmitra, S. Konschuh, C. Ertler, C. Ambrosch-Draxl, and J. Fabian, Phys. Rev. B {\bf 80}, 235431 (2009).

\bibitem{r30} C. K. Safeer, J. Ingla-Ayn\'{e}s, F. Herling, J. H. Garcia, M. Vila, N. Ontoso, M. Reyes Calvo, S. Roche, L. E. Hueso, and F. Casanova, Nano Lett. {\bf 19}, 1074 (2019).

\bibitem{r31} B. Yang, M.-F. Tu, J. Kim, Y. Wu, H. Wang, J. Alicea, R. Wu, M. Bockrath and J. Shi1, 2D Mater. {\bf 3}, 031012 (2016).

\bibitem{rr31} S. Zihlmann, A. W. Cummings, J. H. Garcia, M. Kedves, K. Watanabe, T. Taniguchi, C. Schonenberger, and P. Makk, Phys. Rev. B {\bf 97}, 075434(R) (2018).

\bibitem{rr32} Z. Wang, D. K. Ki, H. Chen, H. Berger, A. H. MacDonald, and A. F. Morpurgo, Nat. Commun. {\bf 6}, 8339 (2015).

\bibitem{st} Jose H. Garcia, Marc Vila, Aron W. Cummings, and Stephan Roche, Chem. Soc. Rev. {\bf 47}, 3359 (2018); A. Mre\'nca-Kolasi\'nska, B. Rzeszotarski, and B. Szafran, Phys. Rev. B {\bf 98}, 045406 (2018).

\bibitem{mt} T. V\"olkl, T. Rockinger, M. Drienovsky, K. Watanabe, T. Taniguchi, D. Weiss, and J. Eroms, Phys. Rev. B {\bf96}, 125405 (2017); B. Yang, E. Molina, J. Kim, D. Barroso, M. Lohmann, Y. Liu, Y. Xu, R. Wu, L. Bartels, K. Watanabe, T. Taniguchi, and Jing Shi, Nano Lett. {\bf 18}, 3580 (2018).

\bibitem{rr33} M. Gmitra, D. Kochan, P. H\"ogl, and J. Fabian, Phys. Rev. B {\bf 93}, 155104 (2016).

\bibitem{rr34} M. Gmitra and J. Fabian, Phys. Rev. B {\bf 92}, 155403 (2015).

\bibitem{r32} M. Gmitra, D. Kochan, and J. Fabian, Phys. Rev. Lett. {\bf 110}, 246602 (2013).

\bibitem{nnn7} Z. Wang, D.-K. Ki, J. Y. Khoo, D. Mauro, H. Berger, L. S. Levitov, and A. F. Morpurgo1, Phys. Rev. X {\bf 6}, 041020 (2016).

\bibitem{r34} M. Charbonneau, K. M. Van Vliet, and P. Vasilopoulos, J. Math. Phys. {\bf 23}, 318
(1982).

\bibitem{r37} D. Xiao, W. Yao, and Q. Niu, Phys. Rev. Lett. {\bf 99}, 236809 (2007).

\bibitem{nnn6} A. Rycerz, J. Tworzydlo, and C. W. J. Beenakker, Nat. Phys. {\bf 3}, 172 (2007).

\bibitem{r35} Z. Li and J. P. Carbotte,  Phys. Rev. B {\bf 86}, 205425 (2012).

\bibitem{r36} V. P. Gusynin, S. G. Sharapov, and J. P. Carbotte, Phys. Rev. Lett. {\bf 96}, 256802 (2006).

\bibitem{r38} V. Vargiamidis and P. Vasilopoulos, J. Appl. Phys. {\bf 116}, 063713 (2014).

\bibitem{r39} D. S. L. Abergel and V. I. Fal\'{}ko, Phys. Rev. B {\bf 75}, 155430 (2007).

\bibitem{r40} E. J. Nicol and J. P. Carbotte, Phys. Rev. B {\bf 77}, 155409 (2008).

\bibitem{r41} T. Stauber, N. M. R. Peres, and F. Guinea, Phys. Rev. B {\bf 76}, 205423 (2007).

\bibitem{r42} K. Nomura and A. H. MacDonald, Phys. Rev. Lett. {\bf 98}, 076602 (2007).

\bibitem{r43} A. A. Patel and S. Mukerjee, Phys. Rev. B {\bf 86}, 075411 (2012).



\end{thebibliography}
\end{document}